\documentclass[lettersize,journal]{IEEEtran}
\usepackage{cite,color}
\usepackage{amsmath,amssymb,amsfonts}
\usepackage{algorithm}
\usepackage{algorithmicx}
\usepackage{algpseudocode}
\usepackage{array}
\usepackage{url}
\usepackage[normalem]{ulem}
\useunder{\uline}{\ul}{}
\usepackage{float}
\usepackage{subfigure,multirow,bm,stfloats}
\usepackage{textcomp}
\usepackage{amsthm}

\usepackage{graphicx}
\usepackage[table]{xcolor}

\usepackage{booktabs} 

\usepackage{pifont}

\usepackage{bbm}
\usepackage[acronym]{glossaries}
\newcommand{\newac}{\newacronym}
\newcommand{\ac}{\gls}

\newcommand{\acpl}{\glspl}

\newac{ips}{IPS}{Indoor Positioning System}
\newac{wlan}{WLAN}{Wireless Local Area Network}
\newac{ap}{AP}{Access Point}
\newac{sc}{SC}{Subspace Clustering}
\newac{uwb}{UWB}{Ultra-Wideband}
\newac{hmm}{HMM}{Hidden Markov Model}
\newac{em}{EM}{Expectation-Maximization}
\newac{lbs}{LBS}{Location-Based Service}
\newac{gpc}{GPC}{Gaussian Process Classification}
\newac{csi}{CSI}{Channel State Information}
\newac{iot}{IoT}{Internet of Things}
\newac{nlos}{NLoS}{Non-Line-of-Sight}
\newac{gpca}{GPCA}{Generalized Principal Component Analysis}
\newac{msl}{MSL}{Multistage Learning}
\newac{mppca}{MPPCA}{Mixture of Probabilistic Principal Component Analysis}
\newac{alc}{ALC}{Agglomerative Lossy Compression}
\newac{ransac}{RANSAC}{Random Sample Consensus}
\newac{fa}{FA}{Factorization-Based Affinity}
\newac{gpcaa}{GPCAA}{GPCA-Based Affinity}
\newac{slbf}{SLBF}{Spectral Local Best-Fit Flats}
\newac{lsa}{LSA}{Local Subspace Affinity}
\newac{llmc}{LLMC}{Locally Linear Manifold Clustering}
\newac{ssc}{SSC}{Sparse Subspace Clustering}
\newac{lrr}{LRR}{Low-Rank Representation}
\newac{scc}{SCC}{Spectral Curvature Clustering}
\newac{mfb}{MFB}{Matrix Factorization-Based}
\newac{mssc}{MSSC}{Multi-View Low-Rank Sparse Subspace Clustering}
\newac{dsc}{DSC}{Deep Subspace Clustering}
\newac{sssc}{SSSC}{Stochastic Sparse Subspace Clustering}
\newac{mdssc}{MDSSC}{Multi-Dimensional Sparse Subspace Clustering}
\newac{nmi}{NMI}{Normalized Mutual Information}
\newac{svm}{SVM}{Support Vector Machine}
\newac{tdoa}{TDoA}{Time Difference of Arrival}
\newac{gmm}{GMM}{Gaussian Mixture Model}
\newac{pca}{PCA}{Principal Component Analysis}
\newac{dnn}{DNN}{Deep Neural Network}
\newac{ann}{ANN}{Artificial Neural Network}
\newac{mlp}{MLP}{Multilayer Perceptron}
\newac{admm}{ADMM}{Alternating Direction Method of Multipliers}
\newac{gan}{GAN}{Generative Adversarial Network}
\newac{rp}{RP}{Reference Position}
\newac{rss}{RSS}{Received Signal Strength}
\newac{wcl}{WCL}{Weighted Central Localization}
\newac{knn}{KNN}{$k$-Nearest Neighbor}
\newac{los}{LoS}{Line-of-Sight}
\newac{toa}{ToA}{Time of Arrival}
\newac{aoa}{AoA}{Angle of Arrival}
\newac{pdf}{PDF}{Probability Density Function}
\newac{cdf}{CDF}{Cumulative Distribution Function}
\newac{rnn}{RNN}{Recurrent Neural Network}
\newac{imu}{IMU}{Inertial Measurement Unit}
\newac{mae}{MAE}{Mean Absolute Error}
\newac{nrmse}{NRMSE}{Normalized Root Mean Square Error}
\newac{rmse}{RMSE}{Root Mean Square Error}
\newac{map}{MAP}{Maximum A Posteriori}
\usepackage{enumitem} 

\usepackage{utfsym} 

\begin{document}

	
	
	
	\title{{\color{black}Survey-Free} Radio Map Construction via HMM-Based Coarse-to-Fine Inference}

	\author{
	Zheng~Xing, 
	Weibing~Zhao,
	Guanghui Zhang,
	Guangjin~Pan,  
	Xuhui~Zhang,
	Jinke~Ren, \\
	Henk~Wymeersch, \textit{Fellow, IEEE},
	Yuan~Wu, \textit{Senior Member, IEEE}, \\
	and~Shuguang~Cui, \textit{Fellow, IEEE}
\thanks{
	Zheng Xing, Xuhui Zhang, Weibing Zhao, Jinke Ren, and Shuguang Cui  are with FNii, The Chinese University of Hong Kong (Shenzhen), Shenzhen, Guangdong 518172, China. (Corresponding author: Zheng Xing. Email: zhengxing@link.cuhk.edu.cn.)}
\thanks{Guanghui Zhang is with School of Computer Science and Technology, Shandong University, Qingdao 266237, China.}
\thanks{Guangjin Pan and Henk Wymeersch are with the Department of Electrical Engineering, Chalmers University of Technology, Gothenburg, Sweden.
}
\thanks{Yuan Wu is with the State Key Lab of Internet of Things for Smart City, University of Macau, Macao SAR, China.
}
}
		
		
		

	
	\maketitle
\begin{abstract} 
	Traditional radio map construction methods mandate labor-intensive data collection and precise location labeling. To address these limitations, we propose a novel survey-free approach for radio map construction that relies solely on unlabeled \ac{rss} measurements, thereby obviating the need for manual site surveys or auxiliary \acpl{imu}. The key idea involves embedding multiple unlabeled RSS sequences into a known indoor layout, specifically targeting corridor-guided environments with a dominant unidirectional pedestrian flow. However, aligning the embedded coordinates with the \ac{rss} collection locations remains challenging due to the random fluctuations inherent in \ac{rss} data.
	To tackle this, we introduce a \ac{hmm}-based Coarse-to-Fine Inference (HCFI) framework. At the coarse level, we employ an HMM-based region label inference algorithm to partition RSS sequences and align the RSS segments with specific physical regions using graph-based inference. At the fine level, we develop an HMM-based location label inference technique to estimate RSS collection coordinates by leveraging \ac{rss} propagation principles while incorporating sequential spatio-temporal mobility probability. Empirical results from an office environment demonstrate that the proposed method achieves a radio map construction \ac{mae} of 8.96 dB. Furthermore, based on the estimated radio map, \ac{knn} localization yields an average positioning error of approximately 3.33 meters, offering a highly viable, survey-free solution for radio map construction under sequential topological assumptions.
\end{abstract}

\glsresetall
\section{Introduction}

Driven by the ubiquitous proliferation of mobile devices such as smartphones, laptops, tablets, and smartwatches, \acpl{lbs} have garnered increasing attention across both academic and industrial sectors \cite{Zaf:J19,pan2025ai,ref7}. Advanced localization techniques, including \ac{toa} \cite{cheng2025hybrid}, \ac{tdoa} \cite{kumar2024joint}, and \ac{aoa} \cite{zheng2025near}, offer precise sub-meter accuracy under \ac{los} conditions. However, these methods necessitate the deployment of specialized, high-precision hardware, which incurs prohibitive costs for both physical maintenance and the continuous calibration required to maintain localization accuracy. In many indoor scenarios, meter-level accuracy is often sufficient, making cost-effectiveness more crucial than achieving sub-meter precision. Consequently, \ac{rss}-based localization has emerged as an attractive alternative due to its inherent low-cost advantages \cite{ref5}. Existing RSS-based algorithms encompass a wide range of model-based, model-free, and data-driven methodologies \cite{bahl2000radar, PraSur:20, lee:J22, Pand:J22, Jin:J20, WanGan:J21, GuoNi:J22, WanUrr:LJ11, Anagnostopoulos2016online, Zaf:J19, ZhaHui:J22, XingChen:J23ar, tao2021aips, Xue2018new, ref11, wang2025radiodiff, jia2025radiomamba, ye2022se, Zhou:J21, Zhu2022intelligent}.

Despite its accessibility, the fundamental bottleneck of RSS-based localization lies in the construction of the underlying radio map (or fingerprint database) \cite{11153005, wang2025radiodiff, jia2025radiomamba, lu2025concealing}, which serves as a critical prerequisite for facilitating LBS applications like indoor navigation and asset tracking \cite{11216398,11162449,10648660,10274422}. To adapt to dynamic environments, advanced deep learning architectures are also increasingly employed to extract robust spatial features \cite{Zhou:J21, wang2017cifi, li2021dafi, prasad2023domain}. Traditionally, constructing these maps relies on exhaustive manual site surveys \cite{jointimplicit}. In this process, human operators traverse the environment carrying specialized equipment to record RSS readings at predetermined \acpl{rp} \cite{du2021crcloc}. While yielding precise ground truth data for point-wise localization \cite{Xie:J23, Tran:J08, Wu:J07, Haj:J22, Wang:J22, Pand:J22, chen2025fas, johnny2025reflection}, these site surveys are inherently labor-intensive, time-consuming, and scale poorly. Furthermore, they require frequent recalibrations to maintain accuracy when environmental changes occur, such as furniture rearrangement, AP displacement, or variations caused by intrinsic hardware heterogeneity \cite{ye2022se,Zaf:J19,ref11,zou2017winips,wu2012will,li2021transloc, fang2015novel}. 

While various spatial interpolation and data augmentation techniques (e.g., matrix factorization, graph representation, meta-learning) have been proposed to reduce the volume of labeled data required \cite{zhu2023protecting, LiyNis:C19, CheKev:C21, wang2023leto, ye2018rmapcs, yin2008learning, tao2025lcwf, ref11, ye2022se, li2021transloc, rai2012zee, xu2021efficient, ShrSag:J22, Weiwu:J23, yassine2022leveraging, ColAnt:J21, FanJic:J21, zhou2025multi, hsiao2023salc, ref17, huang2019online, tao2018novel, wu2017automatic, atia2012dynamic, zou2017winips, wu2015static, sorour2014joint, mu2021intelligent, ChoJeo:J22}, they still fundamentally depend on some degree of location-specific measurement data. To completely obviate the need for site-specific calibration, a variety of crowdsourcing approaches have leveraged mobile users to collect RSS measurements while simultaneously acquiring real-time motion information via \acpl{imu}. This paradigm enables \textit{implicit calibration}, offering advantages in scalability and continuous maintenance. As demonstrated in Table \ref{tab:Literature_method}, systems like Zee \cite{rai2012zee}, UnLoc \cite{wang2012no}, and Leto \cite{wang2023leto}, along with other semi-automated frameworks \cite{wu2014smartphones, li2021wifi, GaoHar:J17, zhang2017unambiguous, jung2015unsupervised, wu2012will, WuYan:J12}, integrate IMU traces (e.g., pedometry, dead reckoning) with Wi-Fi markers or floor plans to rectify radio maps without manual surveys.

\begin{table*}
	\caption{Comparison with location label inference methods without calibration.}
	\label{tab:Literature_method}
	\begin{centering}
		\resizebox{1\textwidth}{!}{
			\begin{tabular}{lcccccccl}
				\hline 
				\multicolumn{1}{l}{Method} & Accelerometer & Gyroscope & Compass & Floor plan & Landmark & AoA &AP location& Technique \tabularnewline
				\hline 
				Zee \cite{rai2012zee} & $\checkmark$ & $\checkmark$ & $\checkmark$ & $\checkmark$ & \scalebox{0.75}{$\usym{1F5F4}$} & \scalebox{0.75}{$\usym{1F5F4}$} &\scalebox{0.75}{$\usym{1F5F4}$} & PDR, PF\tabularnewline
				Unloc \cite{wang2012no} & $\checkmark$ & $\checkmark$ & $\checkmark$ & $\checkmark$ & $\checkmark$& \scalebox{0.75}{$\usym{1F5F4}$}& \scalebox{0.75}{$\usym{1F5F4}$} & PDR\tabularnewline
				LiFS \cite{wu2014smartphones} & $\checkmark$ & \scalebox{0.75}{$\usym{1F5F4}$} & \scalebox{0.75}{$\usym{1F5F4}$} & $\checkmark$ & $\checkmark$ & \scalebox{0.75}{$\usym{1F5F4}$}&\scalebox{0.75}{$\usym{1F5F4}$}  & MDS, graph alignment\tabularnewline
				WiFi-RITA \cite{li2021wifi} & $\checkmark$ & $\checkmark$ & \scalebox{0.75}{$\usym{1F5F4}$} & \scalebox{0.75}{$\usym{1F5F4}$} &$\checkmark$ & \scalebox{0.75}{$\usym{1F5F4}$}& \scalebox{0.75}{$\usym{1F5F4}$} & PDR, trace merging\tabularnewline
				GraphIPS \cite{zhao2020graphips} & $\checkmark$ & $\checkmark$ & $\checkmark$ & \scalebox{0.75}{$\usym{1F5F4}$} & $\checkmark$ & $\checkmark$ &\scalebox{0.75}{$\usym{1F5F4}$} & MDS\tabularnewline
				Leto \cite{wang2023leto} & $\checkmark$ & \scalebox{0.75}{$\usym{1F5F4}$} & \scalebox{0.75}{$\usym{1F5F4}$} & \scalebox{0.75}{$\usym{1F5F4}$} & $\checkmark$ & \scalebox{0.75}{$\usym{1F5F4}$}&\scalebox{0.75}{$\usym{1F5F4}$}  & HMDS, AFD\tabularnewline
				VRLoc~\cite{si2025unsupervised} & \scalebox{0.75}{$\usym{1F5F4}$} & \scalebox{0.75}{$\usym{1F5F4}$} & \scalebox{0.75}{$\usym{1F5F4}$} & $\checkmark$ & \checkmark & \scalebox{0.75}{$\usym{1F5F4}$}&$\scalebox{0.75}{$\usym{1F5F4}$} $ & HMM  \tabularnewline
				Proposed & \scalebox{0.75}{$\usym{1F5F4}$} & \scalebox{0.75}{$\usym{1F5F4}$} & \scalebox{0.75}{$\usym{1F5F4}$} & $\checkmark$ & \scalebox{0.75}{$\usym{1F5F4}$} & \scalebox{0.75}{$\usym{1F5F4}$}&$\checkmark$ & HMM  \tabularnewline
				\hline 
		\end{tabular}}
		\par\end{centering}
\end{table*}

However, a critical limitation of these methods is their strict reliance on client-side motion sensors \cite{rai2012zee,wu2014smartphones,cheng2025hybrid,wang2012no,shen2013walkie}. Implementing real-time localization requires mobile users to be equipped with calibrated IMUs. This poses significant practical challenges: 1) \textit{Hardware Heterogeneity:} Low-cost IMUs in typical smartphones suffer from severe noise and drift, degrading the quality of the ground truth labels. 2) \textit{Availability and Permissions:} Localization software may not always have permission to access IMU data. While our previous work \cite{XingChen:J23ar} introduced a region-level label learning scheme using completely unlabeled RSS measurements, it remains inapplicable to the recovery of precise geographical coordinates.

Specifically, current survey-free methodologies face two primary challenges. First, at the macroscopic level, conventional unsupervised clustering algorithms \cite{ng2002spectral, Bai:J22, xing2024block} (e.g., K-means, spectral clustering) struggle to establish a robust semantic mapping to physical regions due to the noisy, non-convex distribution of RSS data caused by severe multipath effects. Second, at the microscopic level, existing propagation-based tracking models \cite{WanUrr:LJ11, wang2012no, XingChen:J23ar} depend heavily on pre-calibrated, site-specific parameters, making precise coordinate inference formidable without ground-truth location data.

\begin{figure}[t] 
	\centering 
	\includegraphics[width=1\linewidth]{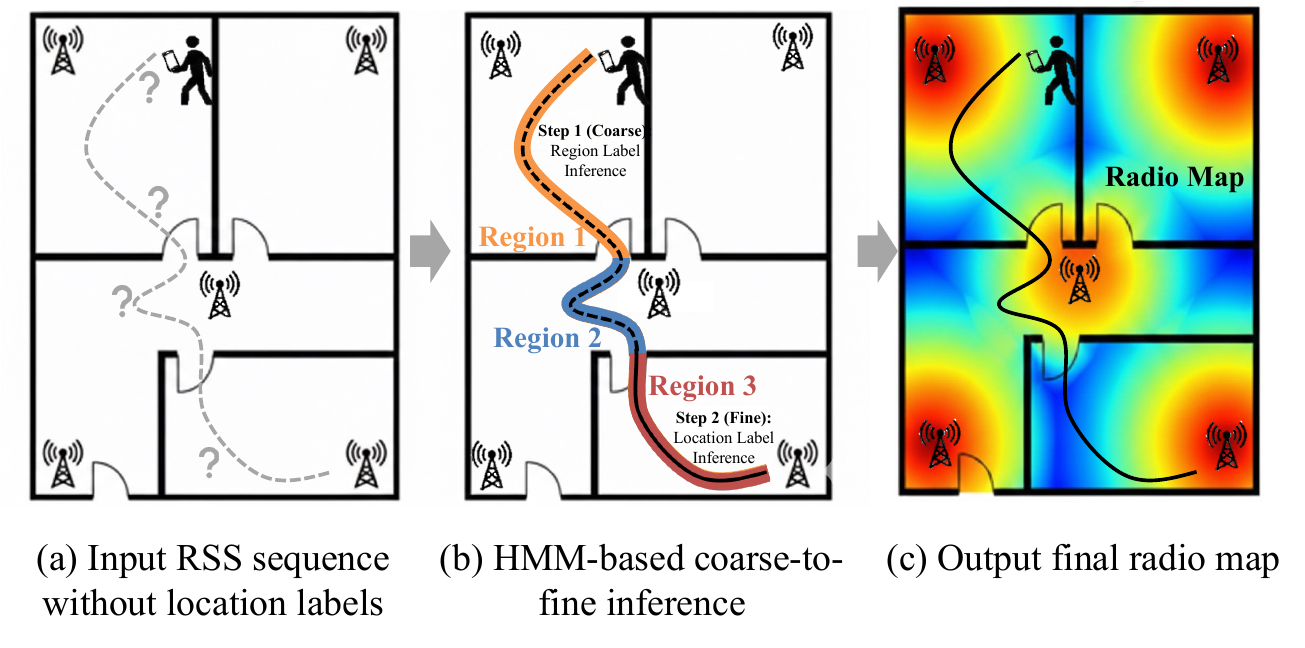} 
	\caption{The proposed framework for survey-free radio map construction. (a) The mobile users collect RSS sequences with unknown locations. (b) The HCFI method applies \ac{hmm}-based coarse-to-fine inference: first grouping RSS data into regional clusters (colored segments), and then estimating the precise trajectory within each region. (c) The final output is the estimated user trajectory and the constructed radio map.} 
	\label{fig:system_structure} 
\end{figure}

To overcome these limitations, we propose the \ac{hmm}-based Coarse-to-Fine Inference (HCFI) framework, as illustrated in Figure \ref{fig:system_structure}. Tailored to corridor-guided environments with a dominant unidirectional flow, HCFI completely eliminates the need for labor-intensive site surveys and auxiliary hardware (e.g., IMUs). By utilizing a known floor plan and established AP positions, our method unsupervisedly recovers data acquisition coordinates from purely unlabeled RSS sequences. It first employs a coarse \ac{hmm}-based inference to robustly cluster RSS data into distinct physical regions. Subsequently, it leverages these regional bounds to estimate the exact collection coordinates and signal propagation parameters through a latent location inference process governed by spatio-temporal constraints. 

Empirical results from an office environment demonstrate that the HCFI-estimated radio map yields an average positioning error of 3.33 meters. While marginally higher than the 2.24-meter error achieved by labor-intensive, manually surveyed maps, our approach offers a highly viable, zero-overhead solution for continuous radio map construction.

To summarize, our main contributions are:
\begin{itemize}[leftmargin=0.5cm]
	\item We introduce a survey-free, IMU-independent framework for constructing radio maps exclusively from unlabeled RSS measurements in corridor-guided environments.
	\item We develop an \ac{hmm}-based region label inference method that integrates subspace learning and \ac{rnn}-based sequence verification to counter signal fluctuations, robustly mapping RSS data to physical regions.
	\item We propose a constrained \ac{hmm}-based location label inference algorithm that iteratively recovers fine-grained spatial coordinates and site-specific propagation parameters by leveraging regional boundaries and the Markovian mobility probability.
\end{itemize}

The remainder of the paper is structured as follows. Section \ref{Sec:system} introduces the system model and problem formulation. Section \ref{Sec:coarse-to-fine} details the proposed HCFI framework. Experimental results are discussed in Section \ref{sec:Experiments}, and the paper is concluded in Section \ref{sec:Conclusion}.
{\color{black}
\section{System Model}
\label{Sec:system}

Consider an indoor environment partitioned into $\mathcal{K}$ distinct, non-overlapping regions. A total of $M$ mobile users navigate this space, adhering to a general, unidirectional topological flow without returning to previously visited regions. Users may selectively skip regions, and their dwelling time within any visited region significantly exceeds the brief inter-region transition time. This structural assumption allows ambiguous transitional RSS samples to be absorbed as clustering noise. By leveraging these natural trajectories, our framework performs implicit calibration, eliminating the need for labor-intensive site surveys.

To clarify the problem scope and system boundaries, Table \ref{tab:assumptions} categorizes the predefined system parameters versus the latent variables that must be inferred through our framework.

\begin{table}[t]
	\caption{Summary of System Assumptions: Known vs. Inferred Variables}
	\label{tab:assumptions}
	\centering
	\resizebox{1\columnwidth}{!}{
		\begin{tabular}{p{0.48\columnwidth} p{0.48\columnwidth}}
			\hline
			\textbf{Known/Given Information} & \textbf{Inferred/Latent Variables} \\
			\hline
			Floor plan layout and reference point sets per region ($\mathcal{V}_k$) & Exact spatial trajectories ($\mathcal{X}_m$) \\
			AP 2D positions ($\mathbf{o}_q$) & Time-slot region labels ($\mathbf{l}_m$) \\
			Total number of regions ($\mathcal{K}$) & Region visit order ($\mathbf{r}_m$) and residence times ($\mathbf{n}_m$) \\
			Unlabeled RSS observation sequences ($\mathcal{Y}_m$) & Region-specific propagation parameters ($\Theta_{k,q}$) \\
			Sequence length per user ($T_m$) & Number of visited regions ($K_m$) \\
			\hline
	\end{tabular}}
\end{table}
Mathematically, the $m$-th user collects a sequence of RSS measurements over $T_m$ discrete time slots. Let $D$ denote the total number of APs, with the known 2D coordinate of the $q$-th AP defined as $\mathbf{o}_q \in \mathbb{R}^2$ for $q \in \{1, 2, \ldots, D\}$. At time slot $t \in \{1, 2, \ldots, T_m\}$, the user's true physical location is $\mathbf{x}_{m,t} \in \mathbb{R}^2$, and the corresponding RSS observation is a $D$-dimensional vector $\mathbf{y}_{m,t} = [y_{m,t,1}, \dots, y_{m,t,D}]^\top \in \mathbb{R}^D$. The complete observation matrix for user $m$ is formulated as $\mathcal{Y}_m = \{\mathbf{y}_{m,1}, \dots, \mathbf{y}_{m,T_m}\}$. This sequence inherently maps to a latent spatial trajectory $\mathcal{X}_m =\{\mathbf{x}_{m,1}, \dots, \mathbf{x}_{m,T_m}\}$ and a discrete sequence of region labels $\mathbf{l}_m = [l_{m,1}, \dots, l_{m,T_m}]$, where $l_{m,t} \in \{1, \dots, \mathcal{K}\}$. The fundamental objective of this paper is to reconstruct $\mathcal{X}_m$ strictly from $\mathcal{Y}_m$ without manual location annotations or inertial sensor data.
}
\begin{figure}[t]
	\centering
	\includegraphics[width=0.5\linewidth]{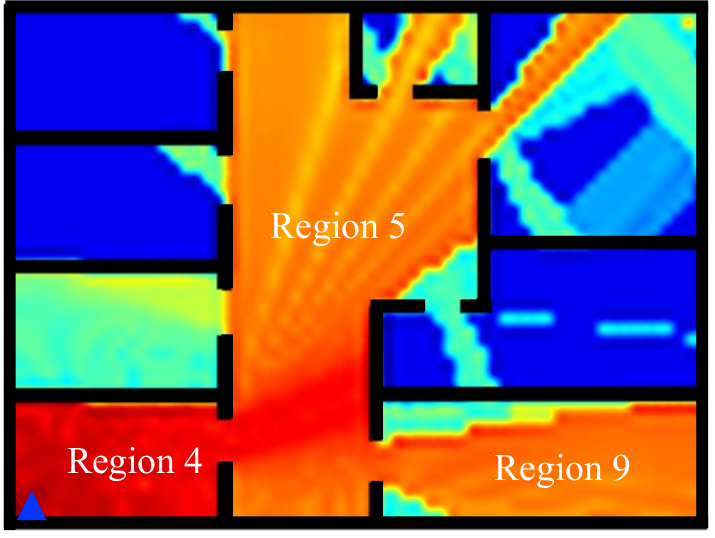} 
	\caption{RSS distribution map for an AP located in Region 4. The AP is denoted by the blue triangle.}
	\label{fig:RSSfit}
\end{figure}

{\color{black}
\subsection{RSS Measurement Model}
\label{subsec:RSS}

While real-world wireless signal propagation is driven by complex physical phenomena, including severe multipath fading, scattering, and diffraction, explicitly modeling these true generative measurement processes through techniques such as 3D ray tracing is computationally prohibitive and often requires highly detailed environmental geometries. Therefore, for the purpose of tractable algorithmic inference, it is necessary to distinguish the true generative process from our analytical model. To characterize the dominant relationship between \ac{rss} and physical distance within the algorithm, we adopt a region-specific log-distance path loss model as our \emph{assumed model}.

As illustrated in Figure~\ref{fig:RSSfit}, a single global model fails to capture the signal distortion caused by physical obstacles such as walls and partitions. Our empirical observations confirm that the assumed path loss model conforms well to the overarching theoretical trend within the host region of the \ac{ap} and extends reliably only to a limited number of neighboring regions where \ac{los} or near-LoS conditions prevail. Specifically, under this assumed model, the \ac{rss} received from the $q$-th AP at time $t$, when the $m$-th user is located in region $k$, is formulated as:
\begin{equation}\label{eq:rss_model}
	y_{m,t,q} = \beta_{k,q} + \alpha_{k,q} \log_{10} \| \mathbf{o}_q - \mathbf{x}_{m,t} \|_2 + \varepsilon_{m,t,q},
\end{equation}
where $\alpha_{k,q}$ and $\beta_{k,q}$ are the path loss exponent and reference power for the $q$-th AP in region $k$. The noise term $\varepsilon_{m,t,q} \sim \mathcal{N}(0, \sigma^2_{k,q})$ is introduced to capture the random fluctuations induced by log-normal shadowing as well as the unmodeled residuals from the complex generative multipath effects. The conditional probability of observing $y_{m,t,q}$ given location $\mathbf{x}_{m,t}$ under the assumed model is thus:
\begin{align}
	&\Pr(y_{m,t,q} \mid \mathbf{x}_{m,t}, \Theta_{k,q}) = \frac{1}{\sqrt{2\pi}\,\sigma_{k,q}} \label{eq:rss_likelihood} \\
	&\quad \times \exp\!\left( -\frac{\left(y_{m,t,q} - \beta_{k,q} - \alpha_{k,q} \log_{10} \| \mathbf{o}_q - \mathbf{x}_{m,t} \|_2\right)^2}{2\sigma^2_{k,q}} \right),\nonumber
\end{align}
where $\Theta_{k,q} = \{\alpha_{k,q}, \beta_{k,q}, \sigma_{k,q}\}$ denotes the unknown propagation parameters for the $q$-th AP in region $k$. Estimating $\Theta_{k,q}$ for each region-AP pair enables a fine-grained characterization of the region-specific propagation conditions. Given the inferred location labels $\{\mathbf{x}_{m,t}\}$ and the region assignments, these assumed propagation parameters $\{\Theta_{k,q}\}$ can be estimated in closed form via least-squares fitting, as detailed in Subsection~\ref{Sec:location-tagging}.
}
{\color{black}
\subsection{UE Mobility Model}
\label{subsec:mobility}
The movement of mobile users within the indoor environment is characterized at two levels. At the \emph{region level}, we model the sequential order in which a user visits different physical regions and the duration of stay within each region. At the \emph{coordinate level}, we model the fine-grained displacement between consecutive time slots, imposing spatio-temporal continuity constraints on the user trajectory. These two levels of modeling correspond to the coarse and fine stages of the proposed framework, respectively.

As mobile users navigate the indoor environment, each user traverses a subset of the $\mathcal{K}$ regions in a sequential manner. While users may physically exhibit arbitrary trajectories (including revisiting rooms) in unconstrained real-world scenarios, our framework specifically targets corridor-guided environments characterized by a dominant unidirectional pedestrian flow. To ensure the tractability of the unsupervised region label inference, we formally assume that users do not revisit previously traversed regions within a single data collection sequence. For the $m$-th user, let $\mathbf{r}_m = [r_{m,1}, r_{m,2}, \ldots, r_{m,K_m}]$ denote the ordered, non-repeating sequence of visited regions, where $r_{m,k} \in \{1, 2, \ldots, \mathcal{K}\}$ and $r_{m,k} \neq r_{m,j}$ for all $k \neq j$. Let $\mathbf{n}_m = [n_{m,1}, n_{m,2}, \ldots, n_{m,K_m}]$ represent the corresponding residence times, where $n_{m,k}$ is the number of time slots spent in region $r_{m,k}$, such that $\sum_{k=1}^{K_m} n_{m,k} = T_m$. The latent region assignment $l_{m,t}$ at time $t$ is then uniquely determined by the pair $\{\mathbf{r}_m, \mathbf{n}_m\}$. For example, if a user visits regions $\mathbf{r}_m = [1, 3, 4]$ with residence times $\mathbf{n}_m = [3, 2, 4]$, the point-wise region label sequence is $\mathbf{l}_m=[1,1,1,3,3,4,4,4,4]$. Both the region visit order $\mathbf{r}_m$ and the residence times $\mathbf{n}_m$ are entirely unknown a priori and must be inferred from the RSS measurements alone. 

Within each region, the movement of the user between consecutive time slots is governed by a spatio-temporal mobility prior. Specifically, let $\delta_t$ denote the time interval between successive slots. In practice, this implicit time resolution depends on the RSS sampling rate of the mobile receiver and is typically on the order of $1$ second. The transition probability from location $\mathbf{x}_{m,t-1}$ to $\mathbf{x}_{m,t}$ is modeled as:
\begin{align}\label{eq:prob-trans}
	&\Pr(\mathbf{x}_{m,t} \mid \mathbf{x}_{m,t-1}) \propto
	\mathbb{I}\!\left\{ \| \mathbf{x}_{m,t} - \mathbf{x}_{m,t-1} \|_2 < v_{\max} \delta_t \right\} \nonumber \\
	&\quad \times \frac{1}{\sqrt{2\pi}\,\sigma_v} \exp\!\left( -\frac{\left( \| \mathbf{x}_{m,t} - \mathbf{x}_{m,t-1} \|_2 / \delta_t - \bar{v} \right)^2}{2\sigma_v^2} \right),
\end{align}
where $\bar{v}$ and $\sigma_v^2$ are the mean and variance of the walking speed of the user, and $v_{\max}$ is the maximum allowable speed. The indicator function enforces a hard constraint on the maximum displacement, while the Gaussian term penalizes deviations from the typical walking speed.
}

\subsection{Problem Formulation}

Based on the RSS measurement model and the user mobility model, we now formulate the radio map construction task as a two-stage inference problem. Since directly estimating the precise coordinates $\{\mathbf{x}_{m,t}\}$ from the raw RSS measurements $\{\mathbf{y}_{m,t}\}$ over the entire floor plan is intractable due to the high-dimensional search space and severe signal fluctuations, we adopt a coarse-to-fine strategy. It is important to clarify that while the physical regions (i.e., their geometric boundaries and the total number $\mathcal{K}$) are pre-decided based on the given floor plan, the actual region labels for individual RSS measurements are strictly latent and unknown a priori. Therefore, the coarse stage first narrows the search space by unsupervisedly inferring which physical region each RSS measurement belongs to. The fine stage then exploits these inferred region assignments to recover the precise coordinates within each region.

\subsubsection{Region Label Inference}

In the coarse stage, the objective is to recover the latent 
region labels $\{l_{m,t}\}$ for all RSS measurements. As 
defined in Subsection~\ref{subsec:mobility}, the region labels are 
uniquely determined by the region visit sequence 
$\mathbf{r}_m$ and the residence times $\mathbf{n}_m$. We 
formulate this as an HMM-based \ac{map} estimation problem, where 
the region assignment $l_{m,t}$ serves as the hidden state 
and the RSS measurement serves as the observation. Applying 
Bayes' rule, the posterior probability of the latent 
variables given the RSS sequence can be factorized as:
\begin{align}\label{eq:Markov}
	p(\mathbf{r}_m, \mathbf{n}_m \mid \mathcal{Y}_m) 
	\propto\; 
	&\prod_{t=1}^{T_m} 
		p(\mathbf{y}_{m,t} \mid l_{m,t})
	\times\; \prod_{k=1}^{K_m} 
		p(n_{m,k} \mid r_{m,k})
	\nonumber \\
	\;\times\;& \prod_{k=1}^{K_m - 1} 
		p(r_{m,k+1} \mid r_{m,k}),
\end{align}
where we assume a uniform prior over the initial region 
$p(r_{m,1})$, which is absorbed into the proportionality 
constant. The factorization follows from the conditional 
independence structure of the HMM: (i) the observation 
likelihood $p(\mathbf{y}_{m,t} \mid l_{m,t})$ assumes that each RSS measurement depends only 
on its current region assignment; (ii) the residence 
duration term $p(n_{m,k}\mid r_{m,k})$ models the number of time slots spent in each 
region independently; and (iii) the transition term $p(r_{m,k+1} \mid r_{m,k})$ captures 
the first-order Markov dependence between consecutively 
visited regions.
{\color{black}
The detailed parameterization of each likelihood component is provided in \eqref{eq:y_l}, \eqref{eq:poisson}, and \eqref{eq:transition} within Subsection~\ref{Sec:region-tagging}.
}

Taking the logarithm of~\eqref{eq:Markov}, the 
\ac{map} estimation is formulated as the following optimization 
problem:
\begin{align}\label{eq:region_opt}
	\underset{\mathbf{r}_m, \mathbf{n}_m}{\text{maximize}} \quad 
	&\sum_{t=1}^{T_m} \log p(\mathbf{y}_{m,t} \mid l_{m,t}) 
	+ \sum_{k=1}^{K_m} \log p(n_{m,k} \mid r_{m,k}) 
	\nonumber \\
	&+ \sum_{k=1}^{K_m - 1} \log p(r_{m,k+1} \mid r_{m,k}) 
	\\
	\text{subject to} \quad 
	&\sum_{k=1}^{K_m} n_{m,k} = T_m, \nonumber \\
	&r_{m,k} \in \{1, 2, \ldots, \mathcal{K}\}, \nonumber \\
	&r_{m,k+1} \neq r_{m,k}, \quad 
	\forall\, k \in \{1, \ldots, K_m - 1\}. \nonumber
\end{align}
The first constraint ensures that the cumulative residence time equals the total sequence length, where the latent number of visited regions $K_m$ is dynamically determined by the length of the inferred sequence (i.e., $K_m = |\mathbf{r}_m|$). The second and third constraints restrict the region labels to valid physical indices and mandate that consecutive segments map to distinct regions.

{\color{black}
Solving \eqref{eq:region_opt} initially yields abstract region clusters lacking explicit spatial semantics. To map these clusters to actual physical regions, we leverage known AP locations as spatial anchors. Specifically, we compute a signal-strength-weighted centroid for each virtual cluster and perform bipartite matching against the known geometric centers of the physical regions. This one-to-one assignment yields the final physical region labels $\{\hat{l}_{m,t}\}$ for the fine stage.
}

\subsubsection{Location Label Inference}

Upon obtaining the region labels $\{\hat{l}_{m,t}\}$ from 
the coarse stage, the fine stage aims to jointly estimate 
the user trajectories $\{\mathbf{x}_{m,t}\}$ and the 
propagation parameters $\{\Theta_{k,q}\}$. We formulate 
this as a sequential \ac{map} inference problem by modeling the trajectory of the user as a Markov chain.

{\color{black}
	
	Assuming a first-order Markov mobility model and the conditional independence of the HMM, the joint probability of the observations and the latent trajectory can be directly factorized. By explicitly conditioning the observation likelihood on the regional propagation parameters $\{\Theta_{k,q}\}$, we obtain:
	\begin{align}\label{eq:joint_recursive}
		&\Pr(\{\mathbf{y}_{m,t}\}_{t=1}^{T_m}, \{\mathbf{x}_{m,t}\}_{t=1}^{T_m} ; \{\Theta_{k,q}\})  \\
		&= p(\mathbf{x}_{m,1}) \prod_{t=1}^{T_m} \Pr(\mathbf{y}_{m,t} \mid \mathbf{x}_{m,t}; \{\Theta_{k,q}\}) \prod_{t=2}^{T_m} \Pr(\mathbf{x}_{m,t} \mid \mathbf{x}_{m,t-1}).\nonumber
	\end{align}

}

As specified in Subsection \ref{subsec:RSS}, the path loss model in \eqref{eq:rss_model} is valid only for a limited number of regions neighboring the AP. To formalize this, we define $\mathcal{Z}_q$ as the region index set encompassing these valid neighboring regions associated with the $q$-th AP. It is important to emphasize that $\mathcal{Z}_q$ is fully known \textit{a priori}. Because the global floor plan (including physical partitions and walls) and the exact AP coordinates are given, $\mathcal{Z}_q$ can be deterministically pre-calculated by evaluating the structural layout to identify which specific regions maintain \ac{los} or near-LoS conditions with the $q$-th AP.

Moreover, assuming a uniform prior $p(\mathbf{x}_{m,1})$ for the initial location, maximizing the joint probability~\eqref{eq:joint_recursive} over all sequences while explicitly masking the invalid AP measurements yields the following optimization problem:
\begin{align}\label{eq:prob-J}
	\underset{\{\mathcal{X}_m\}, \{\Theta_{k,q}\}}{\text{maximize}} \quad
	&\sum_{m=1}^{M} \sum_{t=1}^{T_m} \sum_{q=1}^{D} \mathbb{I}(\hat{l}_{m,t} \in \mathcal{Z}_q) \nonumber\\
    &\times\log \Pr(y_{m,t,q} \mid \mathbf{x}_{m,t}; \Theta_{\hat{l}_{m,t},q}) \nonumber \\
	&+ \sum_{m=1}^{M} \sum_{t=2}^{T_m} \log \Pr(\mathbf{x}_{m,t} \mid \mathbf{x}_{m,t-1}) \\
	\text{subject to} \quad 
	&\mathbf{x}_{m,t} \in \mathcal{V}_{\hat{l}_{m,t}}, \quad \forall\, m, t, \nonumber
\end{align}
where $\mathcal{V}_{\hat{l}_{m,t}} \subset \mathbb{R}^2$ denotes the set of feasible locations within the physical boundary of region $\hat{l}_{m,t}$, and $\mathbb{I}(\cdot)$ is the indicator function. The first term represents the observation likelihood governed by the region-specific path loss model~\eqref{eq:rss_model}, where the indicator function ensures that the $q$-th AP only contributes to the objective if the inferred region $\hat{l}_{m,t}$ of the user falls within its valid set $\mathcal{Z}_q$. The second term $\Pr(\mathbf{x}_{m,t} \mid \mathbf{x}_{m,t-1})$ is the transition probability~\eqref{eq:prob-trans} that encourages spatially smooth trajectories. The constraint $\mathbf{x}_{m,t} \in \mathcal{V}_{\hat{l}_{m,t}}$ confines each location estimate to its assigned region. This constraint, inherited from the coarse stage, is the key mechanism that reduces the search space from the entire floor plan to individual regions, making the joint optimization over $\{\mathcal{X}_m\}$ and $\{\Theta_{k,q}\}$ tractable. The alternating optimization algorithm for solving~\eqref{eq:prob-J} is presented in Subsection~\ref{Sec:location-tagging}.

\section{HMM-Based Coarse-to-Fine Inference Algorithm}
\label{Sec:coarse-to-fine}

\subsection{Region Labels Inference}
\label{Sec:region-tagging}

{\color{black}
\subsubsection{Observation Likelihood Model}

The region label inference problem~\eqref{eq:region_opt} requires evaluating the observation likelihood $p(\mathbf{y}_{m,t} \mid l_{m,t})$ for each RSS measurement. However, in the coarse stage, neither the precise location $\mathbf{x}_{m,t}$ nor the propagation parameters $\{\Theta_{k,q}\}$ are available, precluding the direct use of the physics-based likelihood~\eqref{eq:rss_likelihood}. We therefore construct a statistical surrogate model that characterizes the distribution of RSS measurements within each region. 

It is crucial to emphasize that the true region labels for individual measurements are unknown a priori. Thus, the construction of this surrogate model is inherently iterative. We initialize the region partitions using a global clustering approach (detailed in Subsection~\ref{subsec:em}) and subsequently alternate between evaluating the likelihoods and updating the region-specific parameters via an Expectation-Maximization (EM) framework. This ensures that even if the initial clustering contains errors, the regional assignments are progressively corrected.

In practice, RSS measurements collected by different mobile devices exhibit significant distribution shifts. To mitigate these effects, we pass the raw RSS input $\mathbf{y}_{m,t}$ through a \ac{rnn} to obtain a transformed feature embedding $\tilde{\mathbf{y}}_{m,t}$. The RNN parameters $\Phi$ are jointly learned within the overall EM framework. All subsequent modeling in the coarse stage operates on the transformed embeddings $\tilde{\mathcal{Y}}_m = [\tilde{\mathbf{y}}_{m,1}, \ldots, \tilde{\mathbf{y}}_{m,T_m}]$.

\textit{Subspace model.} 
The $D$-dimensional RSS measurements exhibit severe intra-region variance due to multipath effects. To address this, at each EM iteration, we apply Probabilistic \ac{pca}~\cite{mackiewicz1993principal} to the transformed embeddings based on their current region assignments. This projects the high-dimensional data onto a low-dimensional subspace, suppressing noise while preserving the principal inter-region differences. Specifically, the embedding $\tilde{\mathbf{y}}_{m,t}$ assigned to region $k$ is decomposed as:
\begin{equation}\label{eq:subspace}
	\tilde{\mathbf{y}}_{m,t} = \mathbf{U}_k \boldsymbol{\theta}_{m,t} + \boldsymbol{\mu}_k + \boldsymbol{\epsilon}_{m,t}, \quad \forall (m,t) \in \mathcal{C}_k,
\end{equation}
where $\boldsymbol{\mu}_k \in \mathbb{R}^D$ is the regional mean; $\mathbf{U}_k \in \mathbb{R}^{D \times d_k}$ contains the $d_k$ leading principal directions satisfying $\mathbf{U}_k^\top \mathbf{U}_k = \mathbf{I}$; $\boldsymbol{\theta}_{m,t} \sim \mathcal{N}(\mathbf{0}, \boldsymbol{\Sigma}_k)$ represents the low-dimensional projection coefficients; and $\boldsymbol{\epsilon}_{m,t} \sim \mathcal{N}(\mathbf{0}, s_k^2 \mathbf{I})$ captures the residual noise.
}
\textit{Observation likelihood.} 
From the model in \eqref{eq:subspace}, the likelihood of \( \tilde{\mathbf{y}}_{m,t} \) given its membership in the \( k \)-th subspace is
\begin{align}
	&p_k(\tilde{\mathbf{y}}_{m,t}|l_{m,t}=k,\bm{\tilde{\Theta}})  \label{eq:obs_likelihood}\\
	&=\frac{1}{(2\pi)^{D/2} |\mathbf{C}_k|^{1/2}} \exp\left( -\frac{1}{2} (\tilde{\mathbf{y}}_{m,t} - \bm{\mu}_k)^{\text{T}} \mathbf{C}_k^{-1} (\tilde{\mathbf{y}}_{m,t} - \bm{\mu}_k) \right),\nonumber
\end{align}
where \( \mathbf{C}_k = \mathbf{U}_k \bm{\Sigma}_k \mathbf{U}_k^{\text{T}} + s_k^2 \mathbf{I} \) is the covariance matrix, and \( \bm{\tilde{\Theta}} = \{\mathbf{U}_k, \bm{\Sigma}_k, \bm{\mu}_k, s_k^2\}_{k=1}^{\mathcal{K}} \) collects all parameters. Thus, the general probability is given by
\begin{align}
	\label{eq:y_l}
	p(\tilde{\mathbf{y}}_{m,t}|l_{m,t})=\prod_{k=1}^{\mathcal{K}}p(\tilde{\mathbf{y}}_{m,t}|l_{m,t}=k,\bm{\tilde{\Theta}})^{\mathbb{I}_{\{l_{m,t}=k\}}}.
\end{align}
Although the parameter \( \bm{\tilde{\Theta}} \) is unknown, it is re-estimated at each M-step by maximizing the log-likelihood. Specifically, setting the derivative with respect to \( \bm{\mu}_k \) to zero yields $\hat{\bm{\mu}}_k = \frac{1}{|\mathcal{C}_k|} \sum_{(m,t) \in \mathcal{C}_k} \tilde{\mathbf{y}}_{m,t}$. By defining the weighted covariance matrix \( \mathbf{S}_k = \frac{1}{|\mathcal{C}_k|} \sum_{(m,t) \in \mathcal{C}_k} (\tilde{\mathbf{y}}_{m,t} - \hat{\bm{\mu}}_k)(\tilde{\mathbf{y}}_{m,t} - \hat{\bm{\mu}}_k)^{\text{T}} \), the optimization leads to the eigenvalue problem \( \mathbf{S}_k \mathbf{U}_k = \mathbf{U}_k (\bm{\Sigma}_k + s_k^2 \mathbf{I}) \), providing estimates for $\mathbf{U}_k$, $\bm{\Sigma}_k$, and the noise variance $s_k^2$~\cite{mackiewicz1993principal}.

\subsubsection{Residence Duration Model}

We model the residence time $n_{m,k}$ in region $r_{m,k}$ using a Poisson distribution:
\begin{equation}\label{eq:poisson}
	p(n_{m,k} \mid r_{m,k}) = \frac{\bar{n}_{r_{m,k}}^{n_{m,k}}}{n_{m,k}!} e^{-\bar{n}_{r_{m,k}}},
\end{equation}
where $\bar{n}_{r_{m,k}}$ is the expected number of measurements in region $r_{m,k}$. The Poisson model captures the intuition that larger regions or slower traversal speeds lead to more collected samples, while allowing natural variability.

\subsubsection{Region Transition Prior}

The transition prior $p(r_{m,k+1} \mid r_{m,k})$ encodes structural knowledge about the sequential ordering of regions. Since labels are initially unknown, we first establish a \textit{global} reference ordering $\mathbf{r}^{\text{init}} = [1, 2, \ldots, \mathcal{K}]$ via K-means clustering over all sequences (detailed in Subsection~\ref{subsec:em}). This sequence represents the dominant unidirectional flow of the environment. 
{\color{black}
For instance, if the global corridor flow dictates the order $\mathbf{r}^{\text{init}} = [1, 2, 3, 4]$, User 1 traversing the entire path would exhibit the region sequence $[1, 2, 3, 4]$, while User 2 entering halfway might exhibit a contiguous subsequence $[2, 3, 4]$. Furthermore, as stated in our system model, users may occasionally skip regions (e.g., the sequence of traversed region indices $[1, 3, 4]$) due to rapid movement or sparse signal registrations. Given this reference, the transition prior enforces a unidirectional progression while accommodating such skipped regions:
\begin{equation}\label{eq:transition}
	p(r_{m,k+1} \mid r_{m,k}) \propto
	\begin{cases}
		\displaystyle
		\frac{\bar{n}_{r_{m,k}} + \bar{n}_{r_{m,k+1}}}
		{\sum_{j=r_{m,k}}^{r_{m,k+1}} \bar{n}_j}, 
		& \text{if } r_{m,k+1} > r_{m,k}, \\[6pt]
		0, & \text{otherwise}.
	\end{cases}
\end{equation}
Setting the probability to zero for $r_{m,k+1} \leq r_{m,k}$ strictly enforces the unidirectional, non-revisiting constraint. Importantly, rather than forcibly restricting transitions strictly to consecutive regions, equation~\eqref{eq:transition} naturally handles region skipping. If a user skips intermediate regions (i.e., $r_{m,k+1} - r_{m,k} > 1$), the transition probability is inherently penalized by the expected residence times $\bar{n}_j$ of the skipped regions in the denominator. This elegantly permits occasional skipped regions (such as $[1,3,4]$) while still heavily favoring contiguous, corridor-guided paths.
}

\subsubsection{Alternating Learning and Region Matching}
\label{subsec:em}

The components introduced above involve three groups of unknown parameters: the RNN parameters $\Phi$, the subspace parameters $\tilde{\Theta} = \{\mathbf{U}_k, \boldsymbol{\Sigma}_k, \boldsymbol{\mu}_k, s_k^2\}_{k=1}^{\mathcal{K}}$, and the mean residence times $\Lambda = \{\bar{n}_k\}_{k=1}^{\mathcal{K}}$. We jointly optimize these parameters and infer the latent region labels through a Generalized EM framework.
{\color{black}
\textit{Initialization.} 
The RNN parameters $\Phi^{\text{init}}$ are initialized using standard Xavier initialization. We then apply K-means clustering to the frame-level RSS features across all $M$ sequences to obtain $\mathcal{K}$ initial clusters. Sorting these clusters by their mean temporal positions yields the global transcript $\mathbf{r}^{\text{init}}$. This initial segmentation provides pseudo-labels to initialize the subspace parameters $\tilde{\Theta}^{\text{init}}$ (via \ac{pca}) and the mean residence times $\Lambda^{\text{init}}$. This parameter set constitutes the initial $\theta^{\text{old}} = \{\Phi^{\text{init}}, \tilde{\Theta}^{\text{init}}, \Lambda^{\text{init}}\}$.

\textit{E-step: Viterbi decoding.} 
Given the current model parameters $\theta^{\text{old}}$, obtained from either the initialization or the previous M-step, we compute the \ac{map} estimate of the latent variables, specifically the time-slot region labels $\{l_{m,t}\}$, for each sequence $m$ via Viterbi decoding \cite{forney2005viterbi}. The objective is to find the hidden segment sequence $\{\mathbf{r}_m, \mathbf{n}_m\}$ that maximizes the explicit score function derived from \eqref{eq:region_opt}:
\begin{align}\label{eq:viterbi_score}
	S(\mathbf{r}_m, \mathbf{n}_m) =& \sum_{k=1}^{K_m} \Big[ \log p(n_{m,k} \mid r_{m,k}) \nonumber\\
	&+ \sum_{t=t_{k-1}+1}^{t_k} \log p(\tilde{\mathbf{y}}_{m,t} \mid r_{m,k}) \Big] \nonumber\\
	&+ \sum_{k=1}^{K_m - 1} \log p(r_{m,k+1} \mid r_{m,k}),
\end{align}
where $t_k = \sum_{j=1}^k n_{m,j}$. The Viterbi algorithm efficiently maximizes this score by recursively searching over candidate transition points along the sequence. Once the optimal segments $\{\hat{\mathbf{r}}_m, \hat{\mathbf{n}}_m\}$ are decoded, the time-slot labels $\{\hat{l}_{m,t}\}$ are explicitly assigned.

\textit{M-step: Parameter update.} 
Given the inferred region labels $\{\hat{l}_{m,t}\}$ from the E-step, we update the parameters sequentially.

\emph{(i) RNN parameters $\Phi$:} At each EM iteration, the RNN parameters are updated exclusively via a self-supervised sequence order verification task. Utilizing the label-divided segments from the preceding E-step, we construct a training set $\mathcal{S}$ of concatenated sequences. For positive samples ($y=1$), the segments are concatenated while strictly preserving their true chronological trajectory order. For negative samples ($y=0$), the temporal ordering of these segments is randomly shuffled prior to concatenation. The RNN acts as a binary classifier that processes a given sequence sample $\mathbf{s}$ and outputs a valid-order probability $p_{\Phi}(\mathbf{s})$. The network parameters are updated by minimizing the standard binary cross-entropy loss:
\begin{equation}\label{eq:rnn_loss}
	\mathcal{L}_{\text{RNN}}(\Phi) = -\frac{1}{|\mathcal{S}|} \sum_{(\mathbf{s}, y) \in \mathcal{S}} \Big[ y \log p_{\Phi}(\mathbf{s}) + (1 - y) \log \big(1 - p_{\Phi}(\mathbf{s})\big) \Big].
\end{equation}
By relying purely on this temporal classification task at the M-step, the RNN is forced to learn embeddings that preserve the chronological structure of user transitions. This elegantly prevents feature collapse and ensures the transformed features maintain robust discriminability across adjacent physical regions.

\emph{(ii) Mean residence times $\Lambda$:} For each region $k$, we update the mean count:
\begin{equation}\label{eq:residence_update}
	\bar{n}_k^{\text{new}} = \frac{\sum_{m=1}^{M} \sum_{k'=1}^{K_m} \hat{n}_{m,k'} \cdot \mathbb{I}(\hat{r}_{m,k'} = k)}{\sum_{m=1}^{M} \sum_{k'=1}^{K_m} \mathbb{I}(\hat{r}_{m,k'} = k)}.
\end{equation}

\emph{(iii) Subspace parameters $\tilde{\Theta}$:} After updating $\Phi$, the transformed embeddings are recomputed. The subspace parameters for each region $k$ are re-estimated on the updated clusters $\mathcal{C}_k$ via eigenvalue decomposition. The E-step and M-step alternate until the relative change in the log-posterior falls below $\epsilon_1 = 10^{-3}$. 
}

\textit{Virtual-to-physical region matching.} Upon convergence, the EM procedure yields $\mathcal{K}$ clusters $\{\mathcal{C}_k\}$ indexed by abstract virtual labels. To unambiguously map these to the actual physical regions on the floor plan, we exploit the known AP locations as spatial anchors. We compute a signal-strength-weighted spatial centroid for each virtual cluster:
\begin{equation}\label{eq:cluster_center}
	\hat{\mathbf{p}}_k = \frac{1}{|\mathcal{C}_k|} \sum_{(m,t) \in \mathcal{C}_k} \frac{\sum_{q=1}^{D} w_{m,t,q} \, \mathbf{o}_q}{\sum_{q=1}^{D} w_{m,t,q}},
\end{equation}
where $w_{m,t,q} = 10^{y_{m,t,q}/10}$. Let $\mathcal{V}_k \subseteq \mathbb{R}^2$ denote the set of reference points in the $k$-th physical region, and $\bar{\mathbf{v}}_k = \frac{1}{|\mathcal{V}_k|} \sum_{\mathbf{v} \in \mathcal{V}_k} \mathbf{v}$ its geometric centroid. The optimal one-to-one mapping $\pi = (\pi_1, \ldots, \pi_{\mathcal{K}})$ is obtained by solving the linear assignment problem via the Hungarian algorithm~\cite{kuhn1955hungarian}:
\begin{equation}\label{eq:matching}
	\min_{\pi \in \mathcal{S}_{\mathcal{K}}} \sum_{k=1}^{\mathcal{K}} \| \hat{\mathbf{p}}_k - \bar{\mathbf{v}}_{\pi_k} \|^2.
\end{equation}
After matching, each measurement $(m,t) \in \mathcal{C}_k$ receives the final physical region label $\pi_k$, completing the coarse stage.
The complete pseudocode for this procedure is summarized in Algorithm~\ref{alg:Region-Awareness}.

\begin{algorithm}[t]
	\caption{HMM-Based Region Label Inference and Learning}
	\label{alg:Region-Awareness}
	\begin{algorithmic}[1]
		\State \textbf{Input:} RSS sequences \( \{\mathcal{Y}_m\}_{m=1}^{M} \), AP locations \( \{\mathbf{o}_q\}_{q=1}^{D} \), physical regions \( \{\mathcal{V}_k\}_{k=1}^{\mathcal{K}} \).
		\State \textbf{Initialize:} 
		\State \quad Initialize $\Phi^{\text{init}}$ via Xavier initialization.
		\State \quad Obtain $\mathbf{r}^{\text{init}}$ and pseudo-labels via K-means on RSS features.
		\State \quad Estimate $\tilde{\Theta}^{\text{init}}$ and $\Lambda^{\text{init}}$ using pseudo-labels.
		\State \quad Set $\theta^{\text{old}} = \{\Phi^{\text{init}}, \tilde{\Theta}^{\text{init}}, \Lambda^{\text{init}}\}$ and $\epsilon_1 = 10^{-3}$.
		\Repeat
		\State \quad \textbf{E-step (Viterbi Decoding):}
		\State \quad \quad $\{\hat{\mathbf{r}}_m, \hat{\mathbf{n}}_m\} = \arg\max S(\mathbf{r}_m, \mathbf{n}_m)$ via \eqref{eq:viterbi_score}.
		\State \quad \quad Extract virtual time-slot labels $\{\hat{l}_{m,t}\}$.
		\State \quad \textbf{M-step (Parameter Update):}
		\State \quad \quad $\Phi^{\text{new}} = \arg\min_{\Phi} \mathcal{L}_{\text{RNN}}(\Phi)$ via \eqref{eq:rnn_loss}.
		\State \quad \quad Update $\Lambda^{\text{new}}$ via \eqref{eq:residence_update}.
		\State \quad \quad Update $\tilde{\Theta}^{\text{new}}$ via eigenvalue decomposition.
		\State \quad \textbf{Convergence Check:} 
		\State \quad \quad \textbf{If} $\Delta \log \text{posterior} < \epsilon_1$ \textbf{break}; \textbf{Else} $\theta^{\text{old}} = \theta^{\text{new}}$.
		\Until{convergence}
		\State \textbf{Post-processing (Virtual-to-Physical Matching):}
		\State \quad Compute centroids $\{\hat{\mathbf{p}}_k\}$ via \eqref{eq:cluster_center}.
		\State \quad Obtain mapping $\pi$ by solving bipartite matching \eqref{eq:matching} via Hungarian.
		\State \textbf{Output:} Physical region labels $\{\hat{l}_{m,t}\}$ and region-aligned embeddings.
	\end{algorithmic}
\end{algorithm}

\subsection{Location Label Inference with Region Labels}
\label{Sec:location-tagging}

\begin{algorithm}[tb]
	\caption{HMM-Based Location Label Inference algorithm.}
	\label{alg:Location-Tagging}
	\begin{algorithmic}[1]
		\State {\bfseries Input}: $\{\mathbf{y}_{m,t}\}_{m,t}$, $\{\mathcal{C}_{k}\}$, convergence threshold $\epsilon_2=10^{-3}$
		\State Initialize propagation parameters $\alpha_{k,q}=-20$, $\beta_{k,q}=0$, $\sigma_{k,q}=0$, and user locations $\mathbf{x}_{m,t}$
		\Repeat
		\State Store current trajectory: $\mathbf{x}_{m,t}^{\text{old}} \gets \mathbf{x}_{m,t}$ for all $m, t$
		
		\State Update trajectory $\mathbf{x}_{m,t}$ for all $t\in\{1,2,\dots,T_m\}$ via GA Algorithm
		
		\State Update parameters $\hat{\alpha}_{k,q}, \hat{\beta}_{k,q}$ using \eqref{eq:Theta-solution} and $\hat{\sigma}_{k,q}^{2}$ using \eqref{eq:sigma}
		
		\Until{$\sum_{m=1}^{M}\sum_{t=1}^{T_m} \big\| \mathbf{x}_{m,t} - \mathbf{x}_{m,t}^{\text{old}} \big\|_2 < \epsilon_2$}
		
		\State {\bfseries Output}: location labels $\{\mathbf{x}_{m,t}\}_{t=1}^{T_m}$
	\end{algorithmic}
\end{algorithm}

We observe that there are two blocks of variables in \eqref{eq:prob-J}. The variable $\mathcal{X}_m$ is dependent on $\{\Theta_{k,q}\}$, and $\{\Theta_{k,q}\}$ is dependent on $\mathcal{X}_m$. Therefore, problem \eqref{eq:prob-J} can be divided into two alternating subproblems. 

The first subproblem updates the propagation parameters:
\begin{align}\label{eq:prob-J-1}
	\underset{\{\Theta_{k,q} \mid k \in \mathcal{Z}_q\}_{q=1}^{D}}{\text{maximize}} \;\; &\sum_{m=1}^{M}\sum_{t=1}^{T_m}\sum_{q=1}^{D} \mathbb{I}(\hat{l}_{m,t} \in \mathcal{Z}_q) \nonumber \\
	&\times \log \Pr(y_{m,t,q} \mid \mathbf{x}_{m,t}; \Theta_{\hat{l}_{m,t},q}).
\end{align}

The second subproblem updates the trajectory:
\begin{align}\label{eq:prob-J-2}
	\underset{\mathcal{X}_m}{\text{maximize}} \;\; 
	&f(\{\mathbf{x}_{m,t}\}_{t=1}^{T_m}) = \sum_{t=1}^{T_m} \sum_{q=1}^{D} \mathbb{I}(\hat{l}_{m,t} \in \mathcal{Z}_q) \nonumber \\
	&\times \log\Pr(y_{m,t,q} \mid \mathbf{x}_{m,t}; \Theta_{\hat{l}_{m,t},q}) \nonumber\\
	&+ \sum_{t=2}^{T_m}\log \Pr(\mathbf{x}_{m,t} \mid \mathbf{x}_{m,t-1}),
\end{align}
for $m=1,2,\dots,M$. 
This alternating optimization has significant advantages: firstly, problem \eqref{eq:prob-J-1} has a closed-form solution; secondly, while problem \eqref{eq:prob-J-2} is non-convex, it can be efficiently solved by greedy searching.

{\color{black}
The joint estimation of trajectories and propagation parameters is identifiable under specific geometric and data conditions. First, determining reliable 2D coordinates within any region strictly requires valid RSS measurements from at least three distinct, non-collinear APs (i.e., $|\{q \mid \hat{l}_{m,t} \in \mathcal{Z}_q\}| \ge 3$). Geometries where the valid APs are collinear lead to degeneracies, such as flip ambiguities across the AP axis. Second, the fine-stage solution is highly sensitive to the accuracy of the region labels $\{\hat{l}_{m,t}\}$ provided by the coarse stage. Because the trajectory search space is strictly bounded by $\mathcal{V}_{\hat{l}_{m,t}}$, severe misclassification in the coarse stage would trap the sequence within incorrect physical boundaries. This forces the parameter estimation in \eqref{eq:Theta-solution} to fit erroneous distances, inevitably leading to divergence or convergence to meaningless local optima.
}

\subsubsection{Obtaining the Propagation Model Parameters}
Let $\mathbf{x}_{m,t}$ be the physical coordinate of the user at time $t$, and $\Theta_{k,q}$ denote the localized propagation parameters for the $q$-th AP in the $k$-th region. Assuming that shadowing is independent across sensors, the conditional probability of measuring the valid RSS components at location $\mathbf{x}_{m,t}$ is given by taking the product only over the valid APs:
\begin{align}
	\Pr \left( \mathbf{y}_{m,t} \mid  \mathbf{x}_{m,t}; \{\Theta_{k,q}\} \right)  = \prod_{q: \, k \in \mathcal{Z}_q} \Pr \left( y_{m,t,q} \mid  \mathbf{x}_{m,t}, \Theta_{k,q} \right).
\end{align}

By grouping the terms in \eqref{eq:prob-J-1} according to the physical regions, the objective function can be elegantly rewritten as maximizing $\sum_{q=1}^{D} \sum_{k \in \mathcal{Z}_q} \mathcal{Q}(\Theta_{k,q})$, where:
\begin{align}\label{eq:J}
	\mathcal{Q}(\Theta_{k,q}) &= -\frac{|\mathcal{C}_{k}|}{2}\log(2\pi) - |\mathcal{C}_{k}|\log\sigma_{k,q} \\
	&\quad - \frac{1}{2\sigma_{k,q}^{2}} \left\| \mathbf{B}_{k,q}\begin{bmatrix} \alpha_{k,q} \\ \beta_{k,q} \end{bmatrix} - \mathbf{r}_{k,q} \right\|_2^2. \nonumber 
\end{align}
Here, $\mathcal{C}_{k}$ represents the set of sample indices $(m,t)$ assigned to region $k$. Moreover, $\mathbf{B}_{k,q} \in \mathbb{R}^{|\mathcal{C}_{k}| \times 2}$ is a matrix whose rows correspond to the samples $(m,t) \in \mathcal{C}_{k}$, with each row formulated as $[\log_{10}\|\mathbf{o}_q - \mathbf{x}_{m,t}\|_2, 1]$, and $\mathbf{r}_{k,q} \in \mathbb{R}^{|\mathcal{C}_{k}|}$ is a column vector collecting the corresponding RSS measurements $y_{m,t,q}$. 

The solution to problem \eqref{eq:prob-J-1} is obtained by taking the derivative of $\mathcal{Q}(\Theta_{k,q})$ with respect to $\Theta_{k,q}$ and setting it to zero. For all valid pairs $(k, q)$ where $k \in \mathcal{Z}_q$, this yields the closed-form solution:
\begin{equation}\label{eq:Theta-solution}
	\begin{bmatrix}
		\hat{\alpha}_{k,q}\\
		\hat{\beta}_{k,q}
	\end{bmatrix} = (\mathbf{B}_{k,q}^{\text{T}}\mathbf{B}_{k,q})^{-1}\mathbf{B}_{k,q}^{\text{T}}\mathbf{r}_{k,q},
\end{equation}
and
\begin{equation}\label{eq:sigma}
	\hat{\sigma}_{k,q}^{2} = \frac{1}{|\mathcal{C}_{k}|} \left\| \mathbf{B}_{k,q}\begin{bmatrix}
		\hat{\alpha}_{k,q}\\
		\hat{\beta}_{k,q}
	\end{bmatrix} - \mathbf{r}_{k,q} \right\|_2^2.
\end{equation}

\subsubsection{Trajectory Searching}
Solving the non-convex trajectory inference problem \eqref{eq:prob-J-2} requires searching a high-dimensional continuous space ($\mathbb{R}^{2T_m}$), a task where standard heuristic algorithms typically struggle to converge. To overcome this, we employ a customized Genetic Algorithm (GA) \cite{lambora2019genetic} that drastically prunes the search space by strictly bounding the target coordinates $\mathbf{x}_{m,t}$ within the physical boundaries of their inferred regions $\mathcal{V}_{\hat{l}_{m,t}}$. Compared to conventional search methods, our tailored GA is significantly more robust and efficient because its fitness evaluation and evolutionary operations, including region-constrained random-walk initialization, convex crossover, and clamped mutation, explicitly incorporate both the region-specific path loss likelihoods and the Markovian mobility probability.

\begin{table}[t]
	\caption{Genetic Algorithm (GA) Hyperparameters}
	\label{tab:ga_hyperparameters}
	\centering
	\resizebox{0.8\columnwidth}{!}{
		\begin{tabular}{lc}
			\hline
			\textbf{Hyperparameter} & \textbf{Value} \\
			\hline
			Population Size ($S$) & 100 \\
			Number of Generations ($G$) & 50 \\
			Crossover Rate & 0.8 \\
			Mutation Rate & 0.1 \\
			Selection Method & Tournament (Size = 5) \\
			\hline
	\end{tabular}}
\end{table}
{\color{black}
The specific hyperparameters utilized for the GA are summarized in Table~\ref{tab:ga_hyperparameters}. The algorithm's performance exhibits a clear sensitivity to the population size $S$ and the number of generations $G$. Smaller populations ($S < 50$) or fewer generations ($G < 20$) often lead to premature convergence to local optima, increasing the localization error. Conversely, excessively large values ($S > 200, G > 100$) yield only marginal accuracy improvements while substantially increasing the computational burden. 
}

The complete pseudocode for this procedure is summarized in Algorithm~\ref{alg:Location-Tagging}.
By jointly evaluating the RSS observation reliability and strictly penalizing kinematically anomalous displacements (e.g., exceeding the maximum speed $v_{\max}$ defined in \eqref{eq:prob-trans}), this approach elegantly guarantees physically feasible user trajectories. Regarding the computational complexity, evaluating the fitness of a single trajectory sequence requires computing the region-specific path loss likelihoods across $D$ APs for $T_m$ time slots. Thus, evaluating one individual takes $\mathcal{O}(T_m D)$ operations. For a population of size $S$, one generation requires $\mathcal{O}(S T_m D)$ operations. Since crossover and mutation scale linearly with $S$ and $T_m$, the total time complexity per EM iteration is strictly bounded by $\mathcal{O}(G S T_m D)$, explicitly justifying the linear complexity claim with respect to the sequence length and AP density.


	\section{Experimental Results}
	\label{sec:Experiments}

	In this section, we first describe the experimental setup in Section \ref{subsec:expSet}. Next, we analyze the region label inference performance in Sections \ref{subsec:MapRegConst}. Finally, we evaluate the radio map construction performance and radio map-based localization performance in Sections \ref{subsec:radioMcon} and \ref{subsec:radioMLoc}. 
	\subsection{Experimental Setup}
	\label{subsec:expSet}
	
	\begin{figure}[t]
		\centering
		\includegraphics[width=1\linewidth]{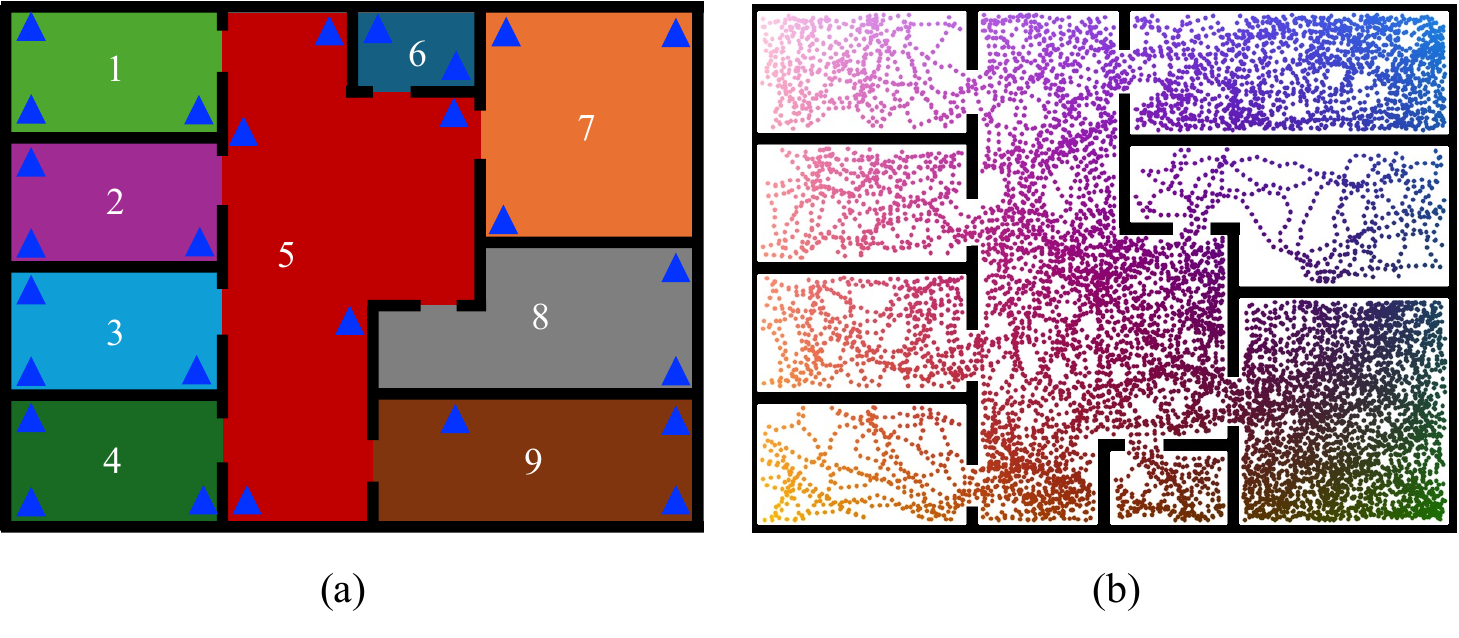} 
		\caption{(a) The indoor environment is partitioned into nine distinct regions, each represented by a different color, with the spatial distribution of 27 APs indicated by blue triangles. (b) Multiple mobile users traverse the indoor space along random trajectories, during which RSS measurements are continuously collected.}
		\label{fig:trajectory}
	\end{figure}
	
The proposed methodology was evaluated in a representative 768 m² office environment, as illustrated by the floor plan and trajectory map in Figure \ref{fig:trajectory}. A total of 27 APs were deployed across nine non-overlapping regions of interest. During the data acquisition phase, $M=4$ mobile participants traversed these target regions along stochastic trajectories to collect \ac{rss} measurements, yielding \ac{rss} sequences with lengths of 12123, 8343, 5335, and 15533 samples, respectively. Subsequently, test datasets I, II, and III were constructed using \ac{rss} data collected independently by three different users at distinct times. These datasets comprise \ac{rss} measurements acquired at 200 randomly distributed locations across different days. For the proposed method, the system parameters were configured as follows: the learning rates were set to $\tilde{\alpha}_1=0.5$ and $\tilde{\alpha}_2=0.1$, the subspace dimension was defined as $d_k=2$, and the maximum speed was set to $v_{\text{max}}=3$ m/s. The number of latent regions was initialized to $\mathcal{K}=9$ to align with the true architectural partitions. 
{\color{black}Furthermore, our framework exhibits strong robustness to minor misspecifications of $\mathcal{K}$ (e.g., $\mathcal{K}=8$ or $10$). When under-specified ($\mathcal{K}=8$), the coarse stage naturally merges adjacent regions with highly correlated signal profiles; when over-specified ($\mathcal{K}=10$), the bipartite matching mechanism absorbs the redundant virtual cluster, and the fine-stage trajectory search gracefully compensates for the boundary discrepancies, resulting in negligible degradation to the final radio map accuracy.}
	
	{\color{black}
	\subsubsection{Baseline Implementations}
	To ensure rigorous reproducibility and address the strict dependencies of different algorithms, we clearly delineate the experimental configurations. All experiments were conducted over 10 independent random seeds, with average results reported. 
	To provide a clear and structured evaluation, the baseline algorithms were categorized according to the specific inference tasks they address:
	
	\textit{1) Baselines for Region Label Inference:}
	These methods are evaluated against our coarse stage to assess clustering performance.
\begin{itemize}
	\item \textit{Unsupervised Methods (SPC, BD-DB, Self-CSC, LSE, SASC, SAGSC, TAGCSC):} Reproduced based on their official open-source implementations or detailed algorithmic descriptions. Inputs were strictly limited to raw RSS sequences. Key hyperparameters were fine-tuned via grid search to ensure optimal competitive performance. Specifically, the subspace dimensions were searched within the range of $d \in \{2, 3, 4, 5\}$, the number of nearest neighbors for graph-based affinity matrices was selected from $k \in \{5, 10, 15, 20\}$, and the regularization trade-off parameters were geometrically tuned over the grid $\{10^{-3}, 10^{-2}, \ldots, 10^2\}$.
	
	\item \textit{Supervised Methods (SVM, KNN, MLP):} Implemented via the \textit{scikit-learn} library. These models were trained utilizing 70\% of the data as labeled training points and optimized via 5-fold cross-validation, serving as an upper-bound reference for region classification. For the specific parameter settings, the SVM utilized an RBF kernel with the penalty parameter $C$ searched over $\{0.1, 1, 10, 100\}$; the KNN neighborhood size was optimally chosen around $K=5$; and the MLP was configured with two hidden layers (128 and 64 neurons), ReLU activation, and trained using the Adam optimizer.
\end{itemize}
	
	\textit{2) Baselines for Location Label Inference and Radio Map Construction:}
	These methods evaluate the fine-stage trajectory recovery and the final radio map accuracy. To ensure a fair comparison with IMU-dependent baselines, synchronized IMU data (accelerometer, gyroscope, and compass) were simultaneously collected alongside the RSS data at a sampling rate of 50 Hz. It is crucial to note that this IMU data suffers from typical smartphone-grade noise and drift. 
\begin{itemize}
	\item \textit{RSS-only Methods (WCL, RRM, VRLoc):} These unsupervised baselines construct radio maps relying solely on the collected raw RSS sequences. To ensure optimal performance, key algorithmic parameters were carefully tuned: for WCL, the weight exponent $g$, which controls the distance decay, was searched over the discrete set $g \in \{1, 2, 3, 4\}$; for RRM, the local manifold regularization parameter was tuned within the grid $\{0.01, 0.1, 1, 10\}$; and for VRLoc, the latent state dimensions and emission variance bounds were aligned with our proposed HMM settings to ensure a fair comparison.
	
	\item \textit{IMU-Assisted Methods (Zee, Unloc, Leto):} Provided with both RSS sequences and the 50 Hz IMU data. Step detection and heading estimation were implemented using standard threshold-based algorithms and complementary filters to replicate the conditions in their original works. Specifically, for Particle Filter (PF)-based tracking (e.g., Zee), the number of particles was set to $N=1000$ with an initial step length variance of $0.05$ m$^2$; for Unloc, the indoor landmark detection confidence threshold was empirically set to 0.8; and for the state-of-the-art Leto, the learning rate for its topological alignment network was set to $10^{-3}$.
\end{itemize}
}
	
	\subsection{Performance of Region Label Inference}
	\label{subsec:MapRegConst}
	
	We evaluate the proposed region label inference method against the classical spectral clustering (SPC) algorithm \cite{ng2002spectral} alongside several state-of-the-art unsupervised baselines (BD-DB \cite{xing2024block}, Self-CSC \cite{Bai:J22}, LSE \cite{Xia:J22}, SASC \cite{kou2023structure}, SAGSC \cite{fettal2023scalable}, and TAGCSC \cite{wei2023adaptive}). 
	
	The comparative performance of these methods is quantitatively assessed using standard clustering metrics \cite{Bai:J22}: clustering accuracy (Acc), normalized mutual information (NMI), F-Score (F1), adjusted rand index (ARI), and precision (Pr). We evaluate the overall region label inference performance using the mean classification error rate $E_{\text{cla}} = \frac{1}{T} \sum_{t=1}^{T} \mathbb{I}\{\hat{l}_{t} \neq l_{t}^{*}\} \times 100\%$.
	Furthermore, to formally quantify the topology recovery capability demanded by the sequential constraints, we define the Topology Accuracy ($\text{Topo-Acc}$) based on the normalized Levenshtein (edit) distance between the inferred region visit sequence $\hat{\mathbf{r}}_m$ and the ground-truth sequence $\mathbf{r}_m^*$:
	\begin{equation}
		\text{Topo-Acc} = \frac{1}{M} \sum_{m=1}^{M} \left( 1 - \frac{\text{Lev}(\hat{\mathbf{r}}_m, \mathbf{r}_m^*)}{\max(|\hat{\mathbf{r}}_m|, |\mathbf{r}_m^*|)} \right) \times 100\%.
	\end{equation}
	
	\begin{table}[t]
		\caption{Region label inference performance of the proposed method and comparisons on the training dataset.}
		\label{tab:Cluster-Perf-Eva} \centering{}%
		\resizebox{1\columnwidth}{!}{
			\begin{tabular}{l|ccccccc}
				\toprule 
				Methods & Acc & NMI & F1 & ARI & Pr & $E_{\text{cla}}$ & Topo-Acc\tabularnewline
				\midrule 
				\multicolumn{8}{c}{Unsupervised Method}\tabularnewline
				\midrule
				SPC~\cite{ng2002spectral} & 31.8 & 32.4 & 34.0 & 27.6 & 46.2 & 28.6 & -- \tabularnewline
				BD-DB \cite{xing2024block} & 32.8 & 28.1 & 36.4 & 34.8 & 33.3 & 26.3 & -- \tabularnewline
				Self-CSC \cite{Bai:J22} & 48.4 & 50.2 & 47.0 & 44.4 & 34.7 & 20.7 & -- \tabularnewline
				LSE \cite{Xia:J22} & 73.4 & 71.5 & 74.6 & 71.5 & 73.0 & 16.4 & -- \tabularnewline
				SASC \cite{kou2023structure} & 60.1 & 57.9 & 54.4 & 59.5 & 60.3 & 21.6 & -- \tabularnewline
				SAGSC \cite{fettal2023scalable} & 63.8 & 62.2 & 65.7 & 67.3 & 48.5 & 18.2 & -- \tabularnewline
				TAGCSC \cite{wei2023adaptive} & 68.3 & 64.3 & 66.8 & 71.2 & 68.3 & 15.3 & -- \tabularnewline
				HCFI w/o RNN & 88.5 & 86.9 & 88.1 & 89.2 & 88.3 & 8.5 & 74.2 \tabularnewline
				HCFI & \textbf{97.8} & \textbf{96.4} & \textbf{97.2} & \textbf{96.6} & \textbf{96.7} & \textbf{2.4} & \textbf{100.0} \tabularnewline
				\midrule
				\multicolumn{8}{c}{Supervised Method}\tabularnewline
				\midrule
				SVM \cite{lin2003study} & -- & -- & -- & -- & -- & 1.6 & -- \tabularnewline
				KNN \cite{zhang2017efficient} & -- & -- & -- & -- & -- & 1.1 & -- \tabularnewline
				MLP \cite{chen2023cyclemlp} & -- & -- & -- & -- & -- & 1.0 & -- \tabularnewline
				\bottomrule
		\end{tabular}}
	\end{table}
	
	\begin{figure}[t]
		\centering
		\includegraphics[width=1\linewidth]{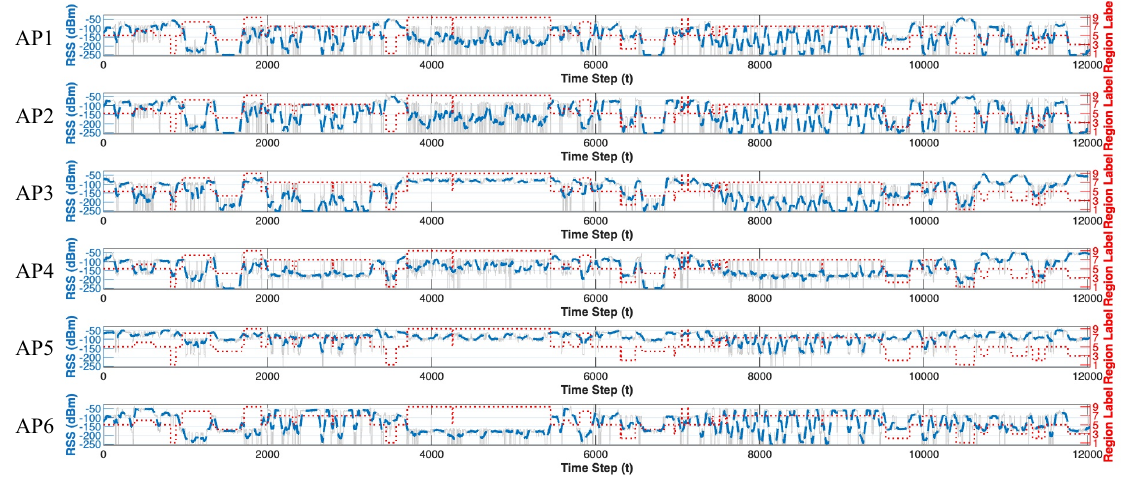} 
		\caption{Illustration of measured RSS from six APs across different labeled regions for a given RSS sequence.}
		\label{fig:RSSplot}
	\end{figure}
	
	\subsubsection{Performance}
	Table~\ref{tab:Cluster-Perf-Eva} summarizes the region label estimation performance on the training dataset. The classical clustering algorithm SPC performs poorly due to its difficulty in handling the significant fluctuations inherent in \ac{rss} measurements as shown in Figure \ref{fig:RSSplot}. Both BD-DB and Self-CSC also exhibit suboptimal clustering performance. LSE demonstrates superior performance among traditional models by employing temporal consistency priors for high-dimensional feature extraction. However, LSE focuses solely on extracting smooth data features via a temporal convolution module, neglecting the temporal assignment consistency required for accurately grouping \ac{rss} measurements in continuous sequences. 
	
	By jointly optimizing the feature embeddings and the HMM priors, our proposed HCFI achieves an exceptional accuracy of 97.8\%, significantly outperforming other unsupervised baseline methods and closely approaching the performance bounds of supervised methods. Furthermore, as formally evaluated by the $\text{Topo-Acc}$ metric, HCFI effectively achieves a 100\% success rate in recovering the correct region-visit sequence topology. The residual 2.4\% classification error arises primarily from minor point-wise assignment deviations precisely at the transition boundaries when a device moves from one region to another.
	
	To explicitly audit the architectural components, we evaluate the ablation variant \textit{HCFI w/o RNN} (i.e., removing the temporal embedding module and the associated binary classification task from the M-step). As shown in Table~\ref{tab:Cluster-Perf-Eva}, removing this temporal embedding module degrades clustering accuracy from 97.8\% to 88.5\%, and severely impairs the topology recovery ($\text{Topo-Acc}$ drops to 74.2\%). This exact consistency between the textual ablation and the table entries corroborates our hypothesis: without the RNN-based sequence verification to smooth short-term anomalies, the network suffers from feature collapse, thereby losing its temporal discriminability across adjacent regions and failing to recover the exact mobility topology.
	
	Finally, Table \ref{tab:Cluster-Perf-Eva} also presents a performance comparison with supervised baselines. While our proposed method applies clustering directly without explicit label guidance, the baseline methods (SVM, KNN, and MLP) are fully supervised, trained on 70\% labeled data. Remarkably, despite this fundamental lack of prior label knowledge, the proposed unsupervised approach achieves a region classification error (2.4\%) that is highly competitive with the fully supervised methods.
	
	\subsubsection{Convergence Analysis}
\begin{figure}[t]
		\centering
		\includegraphics[width=1\linewidth]{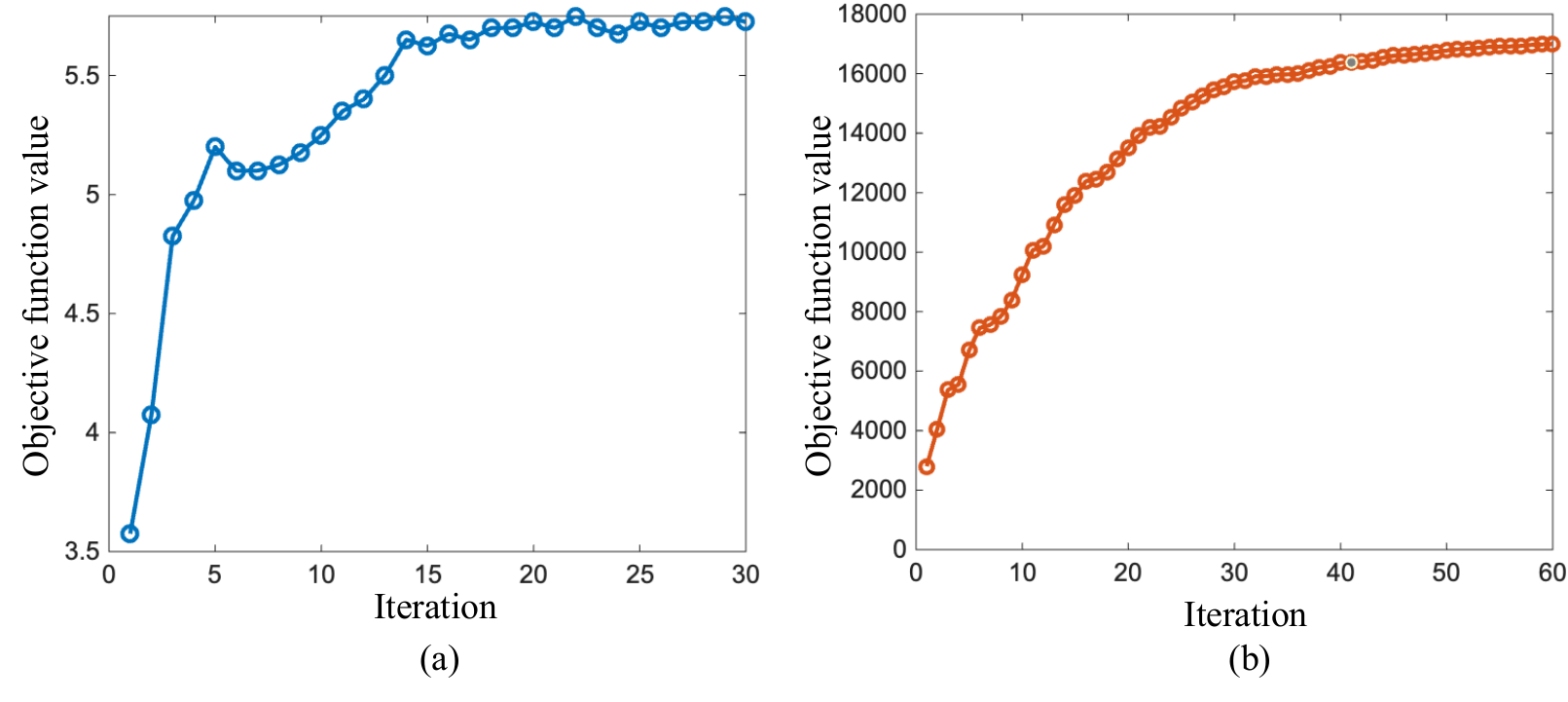} 
		\caption{Convergence analysis of the proposed coarse-to-fine framework. (a) Convergence of the Generalized EM algorithm for region label inference. (b) Convergence of the alternating optimization for location label inference.}
		\label{fig:converge}
	\end{figure}

	To evaluate the convergence of our alternating learning framework in the coarse stage, Fig.~\ref{fig:converge}(a) plots the objective function value of problem \eqref{eq:region_opt}, over the EM iterations on the training dataset. As illustrated in the figure, our Generalized EM algorithm ensures an increasing objective value and converges steadily after approximately 30 iterations.
	
	\subsection{Performance of Radio Map Construction}
	\label{subsec:radioMcon}
	
To evaluate the accuracy of the constructed radio maps and address potential biases in error representation, we employ three standardized metrics: \ac{rmse}, \ac{mae}, and \ac{nrmse}. These metrics are defined as follows: $E_{\text{RMSE}} = \sqrt{\frac{1}{TQ} \sum_{t=1}^{T} \sum_{q=1}^{Q} (e_{t,q} - \hat{e}_{t,q})^2}$, $E_{\text{MAE}} = \frac{1}{TQ} \sum_{t=1}^{T} \sum_{q=1}^{Q} |e_{t,q} - \hat{e}_{t,q}|$, and $E_{\text{NRMSE}} = \frac{E_{\text{RMSE}}}{e_{\max} - e_{\min}} \times 100\%$, where $\hat{e}_{t,q}$ denotes the estimated \ac{rss} at location $\mathbf{x}_t$, and $e_{t,q}$ is the corresponding measured \ac{rss}. In our evaluation, we exclude samples where the \ac{rss} is below $-140$~dBm. $E_{\text{NRMSE}}$ utilizes the global dynamic range ($e_{\max} - e_{\min}$) as the denominator to provide a reliable relative performance measure without the numerical instability associated with individual \ac{rss} scaling. Additionally, we use the average localization error ($E_{\text{loc}}$) to evaluate the accuracy of the \ac{rss}-based location estimation.
	\begin{table}[t]
		\centering
		\caption{Performance of radio map construction on the training dataset.}
		\label{tab:PerLoc_Train}
		\resizebox{\columnwidth}{!}{
			\begin{tabular}{l|rrrr}
				\toprule
				\textbf{Method} & $E_{\text{loc}}$ [m] & $E_{\text{RMSE}}$ [dB] & $E_{\text{MAE}}$ [dB] & $E_{\text{NRMSE}}$ [\%]\\
				\midrule
				WCL \cite{WanUrr:LJ11}   & 15.24 & 27.86 & 20.25 & 21.51 \\
				RRM \cite{XingChen:J23ar}   & 3.14 & 17.62 & 10.65 & 13.60 \\
				Zee \cite{rai2012zee}   & 9.56 & 25.97 & 18.15 & 20.05 \\
				Unloc \cite{wang2012no} & 11.71 & 27.09 & 19.26 & 20.92 \\
				Leto \cite{wang2023leto} & 1.92 & 12.19 & 8.32 & 10.27 \\
				VRLoc~\cite{si2025unsupervised} & 3.65 & 18.79 & 11.60 & 14.51 \\
				HCFI & 2.08 & 15.36 & 8.96  & 11.86 \\
				\bottomrule
			\end{tabular}
		}
	\end{table}
	
	\subsubsection{Performance}
	As shown in Table~\ref{tab:PerLoc_Train}, the proposed HCFI method achieves highly competitive reconstruction accuracy, yielding an $E_{\text{RMSE}}$ of 15.36~dB and an $E_{\text{NRMSE}}$ of 11.86\%. This significantly outperforms classical \ac{rss}-only methods such as WCL, Zee, and Unloc. Furthermore, compared with recent advanced approaches like RRM and VRLoc, the proposed method successfully achieves lower signal construction errors. Although the state-of-the-art Leto method yields slightly lower signal errors ($E_{\text{RMSE}}$ of 12.19~dB), it relies heavily on auxiliary IMU hardware. In contrast, the proposed HCFI framework requires no manual location labels or auxiliary inertial sensors, demonstrating the effectiveness of its region-wise modeling.
	
	In terms of trajectory recovery, the proposed method similarly delivers superior performance among unsupervised, \ac{rss}-only approaches. Table~\ref{tab:PerLoc_Train} indicates that the HCFI method achieves an average location label estimation error ($E_{\text{loc}}$) of just 2.08 meters, outperforming the baselines such as RRM (3.14 m) and VRLoc (3.65 m). While the IMU-assisted Leto method achieves an $E_{\text{loc}}$ of 1.92 meters, our method trails by only 0.16 meters while entirely eliminating hardware dependencies. To validate the necessity of the region-partitioning strategy, we conducted an ablation study where the entire environment was treated as a single undivided region. In this scenario, the proposed method achieves an $E_{\text{loc}}$ of 3.94 meters. While this represents a degradation compared to the 2.08-meter error obtained with multiple-region partitioning, it remains highly competitive, substantiating the robustness of the proposed algorithm.
	
\begin{figure}[t]
		\centering
		\includegraphics[width=1\linewidth]{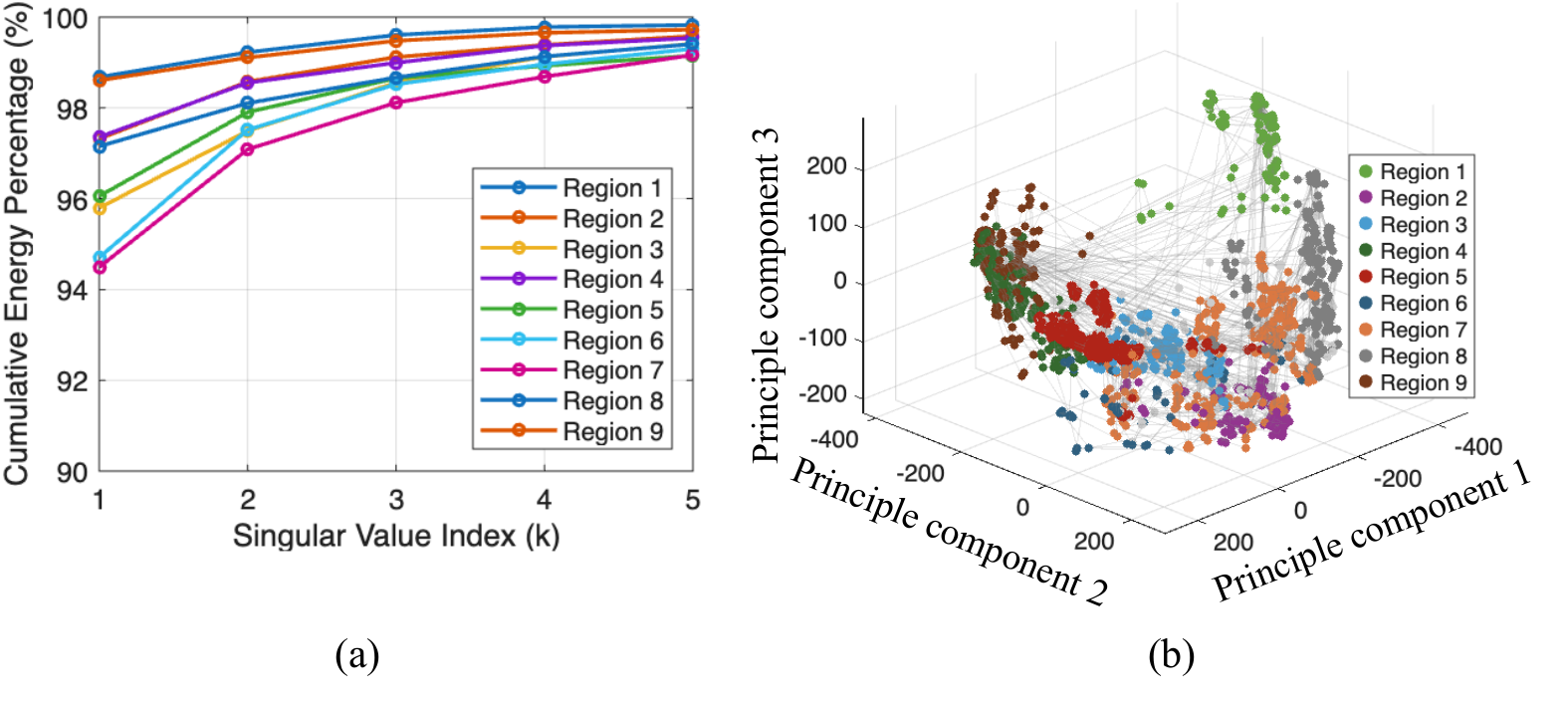} 
		\caption{(a) Singular value spectrum of the RSS measurements with $D=27$ dimension. (b) Spatial distribution of RSS measurements projected into a three-dimensional principle space.}
		\label{fig:svd}
	\end{figure}

	Figure~\ref{fig:svd} validates the effectiveness of the subspace approach. As shown in Figure~\ref{fig:svd}(a), the singular value spectrum reveals that the first few components capture over 96\% of the total energy. Furthermore, Figure~\ref{fig:svd}(b) demonstrates that measurements from different regions form well-separated clusters when projected onto the leading principal components, indicating strong inter-region discriminability.
	
	\begin{figure}[t]
		\centering
		\includegraphics[width=1\columnwidth]{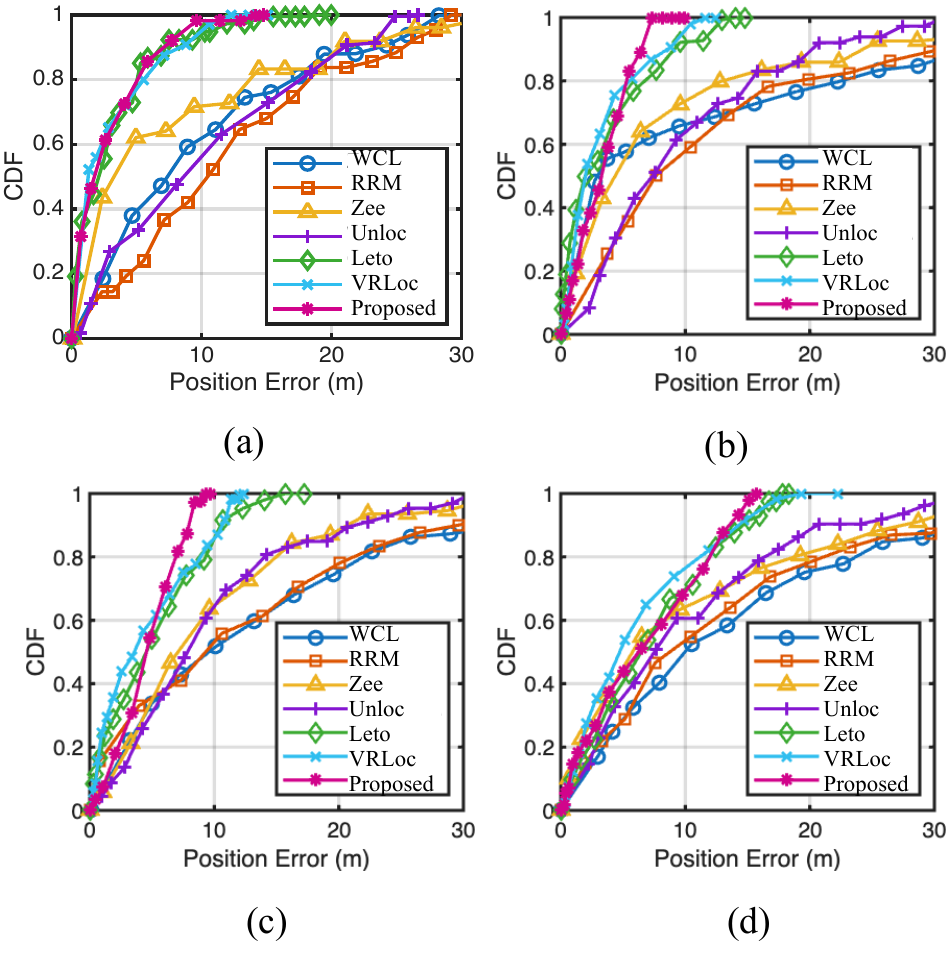}
		\caption{(a) CDF of location label estimation error. (b–d) CDF of localization error for radio map-based positioning on the test datasets. The \ac{knn} algorithm is employed to match the collected \ac{rss} measurements with the constructed radio map.}
		\label{fig:PerLocCDF}
	\end{figure}
	
	Figure~\ref{fig:PerLocCDF}(a) illustrates the \ac{cdf} of the location label inference error. The proposed method exhibits a steep \ac{cdf} curve with the majority of estimation errors falling within 5 meters, demonstrating performance that closely rivals the IMU-based Leto baseline. 
	
	\subsubsection{Convergence Analysis}
	To verify the stability of the fine-stage location label inference, we evaluate the convergence of the alternating optimization process. Fig.~\ref{fig:converge}(b) depicts the value of the joint objective function across successive optimization loops. The monotonically increasing curve demonstrates that the alternating updates efficiently maximize the objective, reaching a stable convergence plateau within approximately 60 iterations.
	
\begin{table}[t]
	\centering
\caption{Performance of radio map-based localization on the test datasets, reported as mean $\pm$ standard deviation over 10 independent random seeds.}
	\label{tab:PerLoc_Test} 
	\resizebox{\columnwidth}{!}{
		\begin{tabular}{l|rrr}
			\toprule 
			\textbf{Method} & \textbf{Test I [m]} & \textbf{Test II [m]} & \textbf{Test III [m]} \\
			\midrule 
			WCL \cite{WanUrr:LJ11}          & 14.33 $\pm$ 1.85 & 14.71 $\pm$ 1.92 & 15.33 $\pm$ 2.05 \\
			RRM \cite{XingChen:J23ar}       & 3.24 $\pm$ 0.45  & 3.56 $\pm$ 0.48  & 3.75 $\pm$ 0.52 \\
			Zee \cite{rai2012zee}           & 11.02 $\pm$ 1.45 & 11.17 $\pm$ 1.50 & 11.74 $\pm$ 1.65 \\
			Unloc \cite{wang2012no}         & 12.07 $\pm$ 1.60 & 12.15 $\pm$ 1.62 & 12.75 $\pm$ 1.75 \\
			Leto \cite{wang2023leto}        & 3.19 $\pm$ 0.40  & 4.55 $\pm$ 0.55  & 4.16 $\pm$ 0.50 \\
			VRLoc~\cite{si2025unsupervised} & 4.22 $\pm$ 0.55  & 4.71 $\pm$ 0.60  & 4.56 $\pm$ 0.58 \\
			Site-Survey                     & 2.24 $\pm$ 0.20  & 2.78 $\pm$ 0.25  & 3.27 $\pm$ 0.30 \\
			HCFI                            & 3.33 $\pm$ 0.25  & 3.61 $\pm$ 0.28  & 3.79 $\pm$ 0.32 \\
			\bottomrule
		\end{tabular}
	}
\end{table}
\subsection{Radio Map-Based Localization}
\label{subsec:radioMLoc}

During the initial phase, RPs were established at spatial intervals of 0.2 m throughout the indoor space. These RPs fulfill two primary functions: first, they provide a uniform grid of candidate locations, allowing the proposed algorithm to identify a sequence of nodes that best align with the measured \ac{rss} sequence; second, they form the spatial basis for constructing the radio maps. Because a stochastic data collection trajectory may not traverse every single RP, we estimate the \ac{rss} values at unvisited RPs using the path loss model derived from our proposed method.

The radio map is constructed using the acquired \ac{rss} measurements alongside their estimated location labels. For any given RP, if multiple \ac{rss} measurements are available, the average value is utilized as the deterministic fingerprint for that location. Conversely, if an RP lacks direct measurements, its \ac{rss} fingerprint is interpolated utilizing established spatial techniques \cite{ye2018rmapcs}. During the online phase, we employ the \ac{knn} algorithm with $K=5$ to match real-time \ac{rss} measurements against the constructed radio map for mobile user localization.

Table~\ref{tab:PerLoc_Test} presents the average localization error (in meters) across the three independent test datasets, comparing radio maps generated by various baseline methods against the proposed HCFI method. The radio map generated by HCFI consistently yields low localization errors, demonstrating superior robustness across diverse testing conditions compared to the state-of-the-art, IMU-assisted Leto baseline. Specifically, while the proposed method trails Leto by a marginal 0.14 meters on Test I, it significantly outperforms Leto by 0.94 meters and 0.37 meters on Test II and Test III, respectively. Furthermore, HCFI consistently yields lower errors than the advanced VRLoc algorithm and maintains performance highly comparable to the RRM method. Notably, the performance gap between HCFI and RRM remains minimal across all sets, narrowing to a mere 0.04 meters on Test III. This is a remarkable feat given that HCFI achieves these results entirely unsupervised, without requiring auxiliary sensor data or manual calibration.

By contrast, traditional \ac{rss}-only approaches such as WCL, Zee, and Unloc fail to maintain accurate spatial representations, consistently exhibiting severe localization errors exceeding 11 meters across all test sets. Furthermore, while the performance of the ground-truth Site-Survey baseline fluctuates by up to 1.03 meters across the different test datasets due to varying data collection conditions, the proposed HCFI method exhibits remarkable consistency. The variation in its average error remains within a narrow 0.46-meter range across the identical datasets, highlighting its stability and strong generalization capabilities.

Figure~\ref{fig:PerLocCDF}(b)--(d) illustrates the \ac{cdf} of the localization errors. The proposed method demonstrates exceptionally high overall reliability across all test scenarios. On test datasets I and II, 100\% of the position errors for the proposed method fall within 10 meters, and on test dataset III, 100\% fall within 15 meters. In contrast, Leto only achieves a 95\% confidence bound at 10 meters for datasets I and II, and a 90\% bound at 15 meters for dataset III. This indicates that HCFI not only provides a highly viable alternative to labor-intensive site surveys but also suppresses extreme localization outliers more effectively across diverse conditions.
\section{Conclusion}
\label{sec:Conclusion}
{\color{black}
In this paper, we investigated the feasibility of survey-free, hardware-independent radio map construction relying exclusively on unlabeled \ac{rss} measurements. Through the development of the \ac{hmm}-based HCFI framework, we draw several key conclusions. First, we demonstrated that the fundamental bottleneck of lacking ground-truth labels and auxiliary inertial sensors such as IMUs can be effectively overcome by mathematically embedding sequential mobility probability into physical indoor boundaries. Second, our empirical results, which achieve a 2.08-meter trajectory error and a 3.33-meter online localization accuracy, prove that unsupervised, region-constrained inference can successfully approach the performance boundaries previously thought to require labor-intensive manual calibration. Ultimately, this work concludes that exploiting the latent spatial correlations within ambient stochastic RF signals provides a highly viable, low-overhead paradigm for sustaining continuous \ac{lbs} in corridor-guided environments.

	Currently, our region label inference methodology relies on a dominant unidirectional flow assumption, which structurally restricts users from complexly revisiting previously traversed regions within a single short-term sequence. While this assumption is highly effective and practical for corridor-guided environments, it inherently limits the framework's applicability in completely unconstrained open spaces. Therefore, an important direction for future work is to relax this topological constraint. We plan to explore the integration of more sophisticated graph matching algorithms or loop-closure detection mechanisms to accommodate arbitrary user trajectories with complex revisiting patterns.
}
		
		
		

	\bibliographystyle{IEEEtran}
	\bibliography{ref}

	\end{document}